\RequirePackage{fix-cm}
\documentclass[preprint]{elsarticle}                     

\usepackage[
    type={CC},
    modifier={by-sa},
    version={4.0},
]{doclicense}

\usepackage{hyperref}

\journal{Brazilian Symposium On Formal Methods (SBMF)}

\usepackage{graphicx}
\usepackage{amsmath}
\usepackage{amssymb}
\usepackage{amsthm}
\usepackage{fancyvrb}
\usepackage{wasysym}
\usepackage{float}
\usepackage{listings}
\usepackage{fvextra} 
\usepackage{alltt}
\usepackage[normalem]{ulem}

\usepackage{xcolor}
\usepackage{hyperref}

\definecolor{Green}{rgb}{0.0, 0.5, 0.0}
\definecolor{mygray}{rgb}{0.5,0.5,0.5}

\usepackage{tikz}
\usetikzlibrary{shapes.geometric}
\usetikzlibrary{patterns}
\usetikzlibrary{decorations.pathreplacing}
\usetikzlibrary{shapes, calc}

\tikzset{
	brace/.style = { decorate, decoration={brace, amplitude=#1pt} },
	mbrace/.style = { decorate, decoration={brace, amplitude=#1pt, mirror} },
	pics/solidPolygon/.style n args={6}{
		code = { %
			\node[regular polygon, regular polygon sides=#1,
			minimum size=#2cm, draw, #3, fill=#4,
			outer sep=0pt] at (#5){{\color{black}#6}};
		}
	},
	pics/hatchedPolygon/.style n args={8}{
		code = { %
			\node[regular polygon, regular polygon sides=#1,
			minimum size=#2cm, draw, #3, pattern = #4, pattern color=#5,
			outer sep=0pt] at (#6){\textcolor{#3}{#7}};
		}
	},
    pics/isoTriang/.style n args={5}{
      code = {
      	\node[isosceles triangle, isosceles triangle apex angle=#3,draw = #1,
      	inner sep=0pt,anchor=lower side,rotate=90, minimum height=#2cm, fill=#4!20] (triangle) at (#5) {};
      }
	},
	pics/isoTriang2/.style n args={6}{
		code = {
			\node[isosceles triangle, isosceles triangle apex angle=#3,draw = #1,
			inner sep=0pt,anchor=lower side,rotate=90, minimum height=#2cm, pattern = #4, pattern color=#5] (triangle) at (#6) {};
		}
	}
}

\newcommand{\nSPolygon}[6]{\pic{solidPolygon = {#1}{#2}{#3}{#4}{#5}{#6}};}

\newcommand{\nHPolygon}[7]{\pic{hatchedPolygon = {#1}{#2}{#3}{#4}{#5}{#6}{#7}};}

\newcommand{\drawterm}[5]{\pic{isoTriang = {#1}{#2}{#3}{#4}{#5}};}
\newcommand{\drawHterm}[6]{\pic{isoTriang2 = {#1}{#2}{#3}{#4}{#5}{#6}};}

\newcommand{\mnt}{\mathrel{\rotatebox[origin=c]{90}{$\twoheadrightarrow$}}}

\newtheorem{definition}{Definition}[section]
\newtheorem{example}{Example}[section]

\newcommand{\noval}{\boldsymbol{\diamondsuit}}

\makeatletter{}\newcommand{\IF}{{\tt IF}\;}
\newcommand{\THEN}{{\tt THEN}\;}
\newcommand{\ELSE}{{\tt ELSE}\;}
\newcommand{\ELSIF}{{\tt ELSIF}\;}
\newcommand{\CASES}{{\tt CASES}\;}
\newcommand{\OF}{{\tt OF}\;}
\newcommand{\LET}{{\tt LET}\;}
\newcommand{\IN}{{\tt IN}\;}

\newcommand{\nat}{\ensuremath{\mathbb{N}}}

\newcommand{\pvso}{\ensuremath{\mathit{pvso}}}

\newcommand{\Expr}{\mathit{expr}}

\newcommand{\Val}{\ensuremath{\mathit{Val}}}
\newcommand{\OPa}{O_1}
\newcommand{\OPb}{O_2}
\newcommand{\False}{\bot}

\newcommand{\vr}{{\tt vr}}
\newcommand{\cnst}{{\tt cnst}}
\newcommand{\opa}{{\tt op1}}
\newcommand{\opb}{{\tt op2}}
\newcommand{\ite}{{\tt ite}}
\newcommand{\rec}{{\tt rec}}

\newcommand{\argp}[1]{\mbox{\tt #1}}

\newcommand{\rhs}{$\mbox{\it rhs}$}
\newcommand{\lhs}{$\mbox{\it lhs}$}

\begin{document}

	\begin{frontmatter}
		
		\title{Formalizing the Dependency Pair Criterion for Innermost Termination}
		\tnotetext[mytitlenote]{Work supported by FAPDF grant 193001369/2016.  \doclicenseThis}
\author{Ariane Alves Almeida$^\dag$\corref{aaalmeida}}
\cortext[aaalmeida]{Author was funded by CAPES with a PhD scholarship.}
\ead{arianealvesalmeida@gmail.com}

\author{Mauricio Ayala-Rinc\'on$^\dag$$^\ddag$\corref{mayalarincon}}
\cortext[mayalarincon]{Corresponding author,    partially   funded   by   CNPq   research   grant   number 307672/2017-4.
}
\ead{ayala@unb.br}

\address{Departments of $^\dag$Computer Science and $^\ddag$Mathematics\\
Universidade de Bras\'ilia}
		
		
		
		
		\begin{abstract}
			Rewriting is a framework for reasoning about functional programming. The dependency pair criterion is a well-known mechanism to analyze termination of term rewriting systems. 
			Functional specifications with an operational semantics based on  evaluation are related, in the rewriting framework, to the innermost reduction relation. This paper presents a PVS formalization of the dependency pair criterion for the innermost reduction relation: a term rewriting system is innermost terminating if and only if it is terminating by the dependency pair criterion.  The paper also discusses the application of this criterion to check termination of functional specifications.		
		\end{abstract}
		
		
	\end{frontmatter}

\section{Introduction}
\label{intro}

Although closely related to the halting problem \cite{Turing36}, and thus undecidable, termination is a relevant property for computational objects.
This property is crucial to state correctness of programs, since it can guarantee that an output will eventually be produced for any input. Even in concurrent and reactive systems, important properties as progress and liveness are related to termination.

It is well-known that  term rewriting systems (TRSs)  are 
an adequate formal framework to reason about functional programs.  
In this context, the dependency pairs (DPs) criterion (\cite{Arts96, ArtsGiesl97, ArtsGiesl98, ArtsGiesl00}), provides a good mechanism to analyze termination. Instead of checking decreasingness of rewrite rules, this criterion aims to check just decreasingness of the fragments of rewrite rules headed by defined symbols. Indeed, a dependency pair consists of the left-hand side (\lhs) of a rewrite rule and a subterm of the right-hand side (\rhs) of the rule headed by  a defined symbol.  Thus, a dependency pair expresses the dependency of a function on calls of any function.  Checking decreasingness over chains of such pairs  corresponds, in a functional specification, to the construction of a \emph{ranking function} that  provides a measure over \emph{data exchanging points} of the program and  that decreases with respect to some well-founded order \cite{Turing1949}. For functional programs, such measures are given over the arguments of each possible (recursive) function call (data exchange point), and it is expected that they decrease after each function call. This is indeed the semantics of termination used in several proof assistants; in particular, in the Prototype Verification Systems (PVS) such \emph{ranking functions} should be provided by the specifier, as part of each recursive definition, and the decreasingness requirements are implemented through the so-called \emph{termination Type Correctness Conditions} (termination TCCs, for short). Termination TCCs are \emph{proof obligations} built by static analysis over the recursive definitions, stating that the measure of the actual parameters of each recursive call strictly decreases regarding the measure of the formal parameters.

Eager evaluation determines the operational semantics of several functional languages, and in particular of the functional language PVS0 specified in PVS for the verification of equivalence between different criteria to automate termination (available as part of the NASA LaRC PVS library at {\color{blue}\url{ https://github.com/nasa/pvslib}}).  The eager evaluation strategy of functional programs corresponds to innermost normalization. Thus to provide formal support to adaptations of the DP criterion over functional programming it is essential to verify the DP criterion for innermost reductions \cite{ArtsGiesl00}.   

\noindent{\bf Main contribution}. This work presents a complete formalization of the DP criterion for innermost reduction.    The formalization extends the PVS library for TRSs (named also {\tt TRS}) that encompasses the basic notions of rewriting as well as some elaborate results (e.g., \cite{Galdino2010}, \cite{Rocha-Oliveira2017}). 
This library includes specifications of terms, positions, substitutions, abstract reduction relations, and term rewriting systems which are adequate for the development of formalizations that remain close to article and textbook proofs, as the one presented in this paper. Although having notions such as noetherianity,  {\tt TRS} did not provide some elements required to fulfill the objective of formalizing the innermost DP criterion. In this sense, this work brings as a minor contribution specifications and formalizations related to the innermost reduction,  non-root reduction and reduction over descendant relations, and as a  major one, the formalization of the equivalence between  the innermost DP criterion and the noetherianity of the innermost reduction relation. 

It is interesting to stress here that the full formalization of the DP criterion for the ordinary rewriting relation is also included in the theory, but since the interesting application is on termination of functional specifications, the focus of this paper is restricted to the innermost reduction case.  The paper also discusses how the DP innermost reduction termination criterion over TRSs is related to the termination of PVS0 functional specifications.

\noindent{\bf Outline}. Section \ref{sec:Basics} gives a brief overview of the basic notions of rewriting and the Dependency Pairs criterion, along with definitions of specific rewriting strategies required in the formalization ahead. Section \ref{sec:Spec} presents the basic elements of the theory {\tt TRS} used in this work along with some additional ones, included by the development of this work, that were required for this formalization.  Section \ref{sec:Necessity} describes the proof that innermost noetherianity  implies termination in the dependency pair criterion, and Section \ref{sec:Sufficiency}  the converse.  Section \ref{sec:RelatedWork} discusses related work, Section \ref{ssec:DPvsFT}  how may be applied this termination criterion to termination of functional programs, and Section \ref{sec:FutureWork} concludes and discusses future work.  The formalization is available as part of the {\tt TRS} library at {\color{blue}\url{http://trs.cic.unb.br}} and also at the NASA PVS library {\color{blue}\url{ https://shemesh.larc.nasa.gov/fm/ftp/larc/PVS-library/}}.

\section{Basic Notions}\label{sec:Basics}

Standard rewriting notation for terms, subterms, positions and substitutions (e.g., \cite{Baader98}), will be used. Given any relation $R$, $R^+$ and $R^*$ denote, respectively, its  transitive and reflexive-transitive closure. The relation $R^*$ between two terms will be referred as \emph{derivation}. For a relation $R$ and element $s$, if there exists $t$ such that $s\;R\; t$ holds, then $s$ is said to be $R$-reducible, otherwise, it is said to be in $R$-normal form, denoted by $nf_{R}(s)$.

A TRS $E$ is a set of rewrite rules that are ordered pairs of terms in $T(\Sigma,V)$, the set of terms freely generated from a countable set of variables $V$ according to a signature $\Sigma$. Whenever the set $E$ is clear from the context, it will be omitted in the notation. 
 Each term $t\in T(\Sigma,V)$ is thus given as a variable or as a function symbol $g$ applied to a tuple of terms of length given by the arity of $g$ according to the signature $\Sigma$.  In order to keep the notation close to the one in the specification, the symbol {\tt f} is not used as a  function symbol, but as the special operator that returns the root function symbol of application terms, which is  automatically created when the datatype for terms is specified. Positions of terms are given as sequences of naturals, as usual: the set of positions of a term $t$, denoted as $Pos(t)$ includes the \emph{root} position that is the empty sequence, denoted as $\lambda$, and if $t$ is an application, say $g(t_1,\ldots,t_n)$, all positions of the form $\{i \pi\;|\; 1\leq i\leq n, \pi\in Pos(t_i)\}$. Given a position $\pi \in Pos(s)$, the subterm of $s$ at position $\pi$ is denoted as  $s|_\pi$. The subterm relation is denoted by $\unrhd$:  $s \unrhd s'$, if there exists $\pi\in Pos(s)$ such that $s'=s|_\pi$. If such given position $\pi$ is such that $\pi \neq \lambda$, $s'$ is called a proper subterm of $s$, which is  denoted as $s \rhd s'$. Notation $s[\pi\leftarrow t]$ is used to denote the term resulting from replacing the subterm at position $\pi(\in Pos(s))$ of $s$ by $t$. 
 
 \begin{example}[Terms, Subterms, TRS, Positions, Reductions, Derivations]\label{ex:trsAck}
 Consider the three rules below conforming a TRS for the Ackermann function, where $s$ and $0$ are the usual constructors for naturals.    
	\[\begin{array}{l}
	a(0,y) \rightarrow s(y) \\
	a(s(x),0) \rightarrow a(x,s(0))\\
	a(s(x),s(y)) \rightarrow a(x, a(s(x),y))
	\end{array}\]
Terms of the form $0, s(0),$ etc are normal forms. Terms of the form $a(0,s^k(0))$ reduce into $s^{k+1}(0)$, and terms of the form $a(s(0),s^{k}(0))$ derive into $s^{k+2}(0)$, for $k>0$, where $s^k$ abbreviates $k$ applications of $s$. The term $a(0,a(s(0),s^k(0)))$ innermost reduces into $a(0, a(0, a(s(0), s^{k-1}(0))))$. Previous innermost reduction happens at position $1$ (the subterm at position $0$ is $0$).
\end{example}

A rewrite rule is denoted by $l \rightarrow r$, and should satisfy the additional restrictions that  $l\notin V$ and that each variable occurring in its right-hand side $r$ also occurs in its left-hand side $l$. Given a TRS $E$, a term $s$ is said to be \emph{reducible at position $\pi \in Pos(s)$} if there exist some rule $l \rightarrow r$, substitution $\sigma$ and term $t$ such that $l\sigma = s|_{\pi}$ and $t = s[\pi \leftarrow r\sigma]$; then $s$ is said to reduce to $t$ at position $\pi$ and is denoted as  $s\overset{\pi}{\rightarrow} t$. If no specific position is given, but there exists some position $\pi \in Pos(s)$ and term $t$ such that $s\overset{\pi}{\rightarrow} t$, $s$ is said to be \emph{reducible},  and whenever $t$ is given, $s$ is said to \emph{reduce} to $t$, denoted as $s \rightarrow_E t$. 

In some specific implementations, such as the one used in this work to deal with chains of \emph{Dependency Pairs}, it is interesting to avoid reductions at  root position of terms. For this, one uses the \emph{non-root reduction} relation, which is denoted by $\overset{>\lambda}{\rightarrow}$,  is induced by a TRS $E$ and relates terms $s$ and $t$ whenever $s\overset{\pi}{\rightarrow} t$ for some $\pi\in Pos(s)$ such that $\pi \neq \lambda$.

 A term $s$ is said to be \emph{innermost reducible at position $\pi \in Pos(s)$} if $nf_{\overset{>\lambda}{\rightarrow}}(s|_{\pi})$ and $s\overset{\pi}{\rightarrow}_{E} t$ for some term $t$; this is denoted as $s\overset{\pi}{\rightarrow}_{i} t$. If no specific position is given, but there exists some position $\pi \in Pos(s)$ and term $t$ such that $s\overset{\pi}{\rightarrow}_{i} t$, $s$ is said to be \emph{innermost reducible}, and whenever $t$ is given, $s$ is said to \emph{innermost reduce} to $t$; this is denoted as $s \rightarrow_{i} t$. Whenever the innermost reduction takes place at a position $\pi \neq \lambda$, one has a so-called \emph{non-root innermost reduction}, denoted by $\overset{>\lambda}{\rightarrow}_{i}$. 

Another important relation in this paper is the descendants of a given term through a given relation. The \emph{reduction relation restricted to (descendants of) a term $t$} is induced by pairs of terms $u,v$ derived from $t$, that is $t \rightarrow^* u$ and $t \rightarrow^* v$, and such that $u \rightarrow v$. The notation used is $\underset{t}{\rightarrow}$. For pairs of terms that are descendants of $t$ and  related one with the other by a reduction at specific position $\pi$, the notation  $\overset{\pi}{\underset{t}{\rightarrow}}$ is used. Analogous notation applies to innermost and non-root reductions. Also, regarding specific terms, a term $s$ is (innermost) terminating if no infinite (innermost) derivation starts with it. If the term is not terminating, the notation $\uparrow$ (or $\uparrow_{i}$) is used. Whenever a term is not terminating, but all its proper subterms are, one says the term is \emph{minimal non-terminating} (\emph{mnt} for short, denoted by $\mnt$), and for innermost termination one says \emph{minimal innermost non-terminating} (\emph{mint} for short, denoted by $\mnt_{i}$).

The termination analysis for rewriting systems aims to verify the non existence of infinite reduction steps (derivations) for every term over which the reduction relation is applied. In order to do this, the DP technique, proposed in \cite{ArtsGiesl97}, analyzes the possible reductions in a term resulting from a previous reduction, i.e., those that can arise from defined symbols on the \rhs's of rules. Thus, it analyzes the  \emph{defined symbols} of a TRS $E$, i.e., the set given by $D_E = \{g \;|\; \exists( l \rightarrow r \in E): {\tt f}(l) = g\}$.

\begin{definition}[Dependency Pairs]\label{def:depPairs}
	Let $E$ be a TRS. The set of \emph{Dependency Pairs} for $E$ is given as 	\[DP(E) = \{\langle l,t \rangle \;|\; l\rightarrow r \in E \;\land\; r \unrhd t \;\land\; {\tt f}(t) \in D_E \}\] 
\end{definition}

\begin{example}[Dependency Pairs]\label{ex:dpAck} The DPs for the TRS  in Example \ref{ex:trsAck} are given below.
	\[\begin{array}{l}
	 \langle a(s(x),0), a(x,s(0))\rangle        \mbox{ from the second rule}\\[1mm]
	 \langle a(s(x),s(y)), a(x,a(s(x),y))\rangle \mbox{ from the third rule at root position}\\[1mm]
	\langle a(s(x),s(y)), a(s(x),y)\rangle   \mbox{ from the third rule at position $1$ of the \rhs}\\[1mm]
\end{array}\]
	\end{example}

Standard definitions of DPs  substitute defined symbols by new {\em tuple symbols} to avoid (innermost) reductions at root positions, which is required for the analysis of termination. Using such tuple symbols (or marked defined symbols) is convenient when using polynomial interpretations since it allows given different interpretations to the defined symbols and their associated tuple symbols (e.g., \cite{ArtsGiesl00}, \cite{ThiemannGiesl03}). 
For the main purpose of this work (that is the formalization of the innermost DP criterion),  and for relating the DP criterion with other termination criteria (available in the PVS theory PVS0), the flexibility allowed by tuple symbols would not be required.  In the current formalization, instead of extending the language with such tuple symbols, DPs are built with unmarked symbols of the original signature and reductions at root position are avoided through the restriction to non-root (innermost) derivations. This choice will be made clearer in Section \ref{sec:Spec}. The advantage of our approach is that in this manner, dealing with new reduction relations over the extended signature is not required.

Each dependency pair represents the possibility of a future reduction after one (innermost) reduction step. However, distinct rewriting redexes can appear in terms after (possibly) several (innermost) reduction steps, which can also give rise to another possible reduction, producing a \emph{Dependency Chain}.

\begin{definition}[Dependency Chain] \label{def:DepChain}
	A \emph{dependency chain} for a TRS $E$, $E$-chain, is a finite or infinite sequence of dependency pairs $\langle s_1, t_1\rangle,\langle s_2, t_2\rangle\ldots$ for which  there exists a substitution $\sigma$ such that $t_i\sigma \overset{>\lambda}{\rightarrow}^* s_{i+1}\sigma$, for every $i$ below the length of the sequence,  after renaming the variables of pairs with disjoint new variables.
\end{definition}

\begin{example}[Dependency Chain]   A dependency chain built using the second DP in the Example \ref{ex:dpAck} is given by: 
\[ \langle a(s(x),s(y)), a(x,a(s(x),y))\rangle,  \langle a(s(x),s(y)), a(x,a(s(x),y))\rangle \]
since $a(s(0),a(s^2(0),0)) \rightarrow^* a(s(0),s(a(s(0),0)))$.
\end{example}

Similarly, the notion of \emph{Innermost Dependency Chain} is given:

\begin{definition}[Innermost Dependency Chain] \label{def:inDepChain}
	An \emph{innermost dependency chain} to a TRS $E$, $E$-in-chain, is a finite or infinite sequence of dependency pairs $\langle s_1, t_1\rangle\langle s_2, t_2\rangle\ldots$ for which there exists a substitution $\sigma$ such that, for every $i$ below the length of the sequence, $t_i\sigma \overset{>\lambda}{\rightarrow}_{i}^* s_{i+1}\sigma$ and $nf_{\overset{>\lambda}{\rightarrow}}(s_i)$, after renaming the variables of pairs with disjoint new variables.
\end{definition}

	Termination is then defined as the absence of infinite (innermost) dependency chains (cf., Theorems  3.2 and 4 of \cite{ArtsGiesl97}).
  
\section{Specification}\label{sec:Spec}

This paper presents an extension of the PVS term rewriting library {\tt TRS}. This library is a development that already contains the basic elements of abstract reduction systems and TRS, such as reducibility, confluence and noetherianity regarding a given relation, notions of subterms and replacement, etc. Furthermore, this theory embraces several elaborate formalizations regarding such systems, such as confluence of abstract reduction systems (see \cite{Galdino08}), the Critical Pair Theorem (see \cite{Galdino2010}) and orthogonal TRSs and their confluence (see \cite{Rocha-Oliveira2017}).

Terms in the theory {\tt TRS} are specified as a datatype with three parameters: nonempty types for variables and function symbols, and the arity function of these symbols.  Terms are either variables or applications built as function symbols with a sequence of terms of length equal to its arity. The predicate {\tt app?} holds for application terms and, as previously mentioned, the operator {\tt f} extracts the root function symbol of an application.  

  The theory {\tt rewrite\_rules.pvs} specifies rewrite rules (as pairs of terms, restricted as usual) and the notion of a set of defined symbols for a set of rewrite rules $E$ (i.e., $D_E$) given as predicate {\tt defined?} in Specification \ref{spec:defsymbol}. 

	\hspace{-4mm}\begin{minipage}{.98\textwidth}\begin{footnotesize}
	\begin{lstlisting}[label = {spec:defsymbol}, caption = {Predicate for defined symbols.}, mathescape, frame = single]
  defined?$(E)(d:symbol)$: bool = 
    $\exists(e \in E)$: f$(\lhs(e)) = d $
	\end{lstlisting}
\end{footnotesize}
\end{minipage}

Basic elements and results were imported in this formalization, such as aforementioned terms, rules and predicates to represent pertinence of positions of a term ({\tt positonsOF} in theory {\tt positions.pvs}), functions to provide subterm of specific position ({\tt subtermOF} in theory {\tt subterm.pvs}), the replacement operation ({\tt replaceTerm} in theory {\tt replacement.pvs}) and so on. However, specification of some general definitions regarding TRS's required to specify DPs and formalization of several properties were missing and filled in as part of this work. Some of these new basic notions and results were included either in existing theories,  such as the notion of non-root reduction ($\overset{>\lambda}{\rightarrow}$) specified in theory {\tt reduction.pvs}, or in new complementary basic theories such as {\tt innermost\_reduction.pvs} and {\tt restricted\_reduction.pvs}, where the relations  $\rightarrow_{i}$ and $\underset{t}{\rightarrow}$ are found. 

Furthermore, the new basic definitions are, mostly, specializations of previously ones, such as the notions presented by predicates {\tt reduction\_fix?} and {\tt reduction?} (see Specification \ref{spec:fixedReduction}), which respectively specify the predicates for relations $\overset{\pi}{\rightarrow}$ and $\rightarrow$ (in theory {\tt reduction.pvs}). Notice that such relations are specified as predicates over pairs of terms in a Curryfied way, a discipline followed through the whole {\tt TRS} library that allows one to rely, for instance, on parameterizable definitions and properties provided for arbitrary abstract reductions systems, such as closures of relations (in theory {\tt relations\_closure.pvs}), reducibitity and normalization (in theory {\tt ars\_terminology.pvs}), noetherianity (in theory {\tt noetherian.pvs}), etc. 

\hspace{-4mm}\begin{minipage}{.98\textwidth}\begin{footnotesize}
	\begin{lstlisting}[label = {spec:fixedReduction}, caption={Predicates for $\overset{\pi}{\rightarrow}$ and $\rightarrow$ relations.}, mathescape, frame = single]
  reduction_fix?$(E)(s, t: $term$ , \pi \in Pos(s))$: bool =
    $\exists(e \in E, \sigma)$:
      $s|_{\pi} =\, \lhs(e)\sigma \land t = s[\pi \leftarrow \rhs(e)\sigma]$

  reduction?$(E)(s, t: $term$ )$: bool =
    $\exists( \pi \in Pos(s))$:
      reduction_fix?$(E)(s, t, \pi)$
     \end{lstlisting}
 \end{footnotesize}
\end{minipage}

The newly required relations are found in theory {\tt innermost\_reduction.pvs}, and are specified as {\tt non\_root\_reduction?} ($\overset{>\lambda}{\rightarrow}$), {\tt is\_nr\_normal\_form?} ($nf_{\overset{>\lambda}{\rightarrow}}$), {\tt innermost\_reduction\_fix?} ($\overset{\pi}{\rightarrow}_{i}$), {\tt innermost\_reduction?} ($\rightarrow_{i}$) and finally {\tt non\_root\_innermost\_reduction?} ($\overset{>\lambda}{\rightarrow}_{i}$).



 
\hspace{-4mm}\begin{minipage}{.98\textwidth}\begin{footnotesize}
	\begin{lstlisting}[label = {spec:nrreduction}, caption={Predicates for the $\overset{>\lambda}{\rightarrow}$, $nf_{\overset{>\lambda}{\rightarrow}}$, $\overset{\pi}{\rightarrow}_{i}$, $\rightarrow_{i}$ and $\overset{>\lambda}{\rightarrow}_{i}$ relations.}, mathescape, frame = single, belowcaptionskip = 8pt]      
  non_root_reduction?$(E)(s,t)$: bool =  
    $\exists(\pi \in Pos(s) |\; \pi \neq \lambda)$: 
      reduction_fix?$(E)(s,t,\pi)$
      
  
  is_nr_normal_form?$(E)(s)$: bool =
    $\forall(\pi \in Pos(s) |\; \pi \neq \lambda)$:
      is_normal_form?$($reduction?$(E))(s|\;_{\pi})$
 
  innermost_reduction_fix?$(E)(s,t,(\pi \in Pos(s)))$: bool = 
    is_nr_normal_form?$(E)(s|\;\pi)) \;\land$ reduction_fix?$(E)(s,t,\pi)$
  
  innermost_reduction?$(E)(s,t)$: bool =  
    $\exists(\pi \in Pos(s))$:
      innermost_reduction_fix?$(E)(s,t,\pi)$
      
  non_root_innermost_reduction?$(E)(s,t)$: bool =  
    $\exists(\pi \in Pos(s) |\; \pi \neq \lambda)$: 
      is_nr_normal_form?$(E)(s|_{\pi}) \;\land$ reduction_fix?$(E)(s,t,\pi)$
 \end{lstlisting}
\end{footnotesize}
\end{minipage}   
    
    The notion of $\underset{s}{\rightarrow}$ is given in Specification \ref{spec:rest} as {\tt rest?} for any binary relation $R$ in theory {\tt restricted\_reduction.pvs}. A specialization of restricted relations for term rewriting is given by {\tt arg\_rest?}, allowing to fix the argument where innermost reductions can take place between given descendants of a term $s$ (i.e., relation $\overset{\pi}{\underset{t}{\rightarrow}}$), which is specified in theory {\tt innermost\_reduction.pvs}. The function {\tt first($\pi$)} returns the first element of the sequence of naturals given by the position $\pi$. 
    
\hspace{-4mm}\begin{minipage}{.98\textwidth}\begin{footnotesize}
  \begin{lstlisting}[label = {spec:rest}, caption = {Predicates for the $\underset{s}{\rightarrow}$ and $\overset{\pi}{\underset{t}{\rightarrow}}$ relations.}, mathescape, frame = single, belowcaptionskip = 8pt]      
      
  rest?$(R, s)(u,v) $: bool = 
    $(s \, R^* \, u) \land (s \, R^*\, v )\land (u\,  R \, v)$
    
  arg_rest?$(E)(s)(k)(u,v)$: bool = 
    rest?$(\overset{>\lambda}{\rightarrow}_{i},s)(u,v) \quad \land$
    $\exists(\pi \in Pos(s) |\; \pi \neq \lambda)$: 
      first$(\pi) = k$ $\land$ 
      innermost_reduction_fix?$(E)(u,v,\pi)$  
  
\end{lstlisting}
\end{footnotesize}
\end{minipage}

Previously mentioned discipline of Curryfication and modularity of {\tt TRS}  that allows generic application of rewriting predicates and their properties over general rewriting relations is followed. For instance, in the specification of  {\tt arg\_rest?} (Specification \ref{spec:rest}), the predicate {\tt rest?} receives as parameter the relation  $\overset{>\lambda}{\rightarrow}_{i}$, satisfying {\tt non\_root\_innermost\_reduction?$(E)$}.

 In theory {\tt dependency\_pairs.pvs}  the notion of DP  and its  termination criterion are specified. As previously mentioned,  instead of extending the language with tuple symbols,  DPs are specified with the same language of the given signature, and thus  DPs chained through  non-root (innermost) reduction.

\hspace{-4mm}\begin{minipage}{.98\textwidth}\begin{footnotesize}
	\begin{lstlisting}[label = {spec:DPstd}, caption = {Predicate for Dependency Pairs as pairs of terms.}, mathescape, frame = single]
  dep_pair?$(E)(s,t)$: bool =
    app?$(t) \; \land$ defined?$(E)($f$(t)) \; \land$       
    $\exists(e \in E)$:$\lhs(e) = s \; \land  (\exists (\pi \in Pos(\rhs(e))): \; \rhs(e)|_{\pi}=t)$
	\end{lstlisting}
\end{footnotesize}
\end{minipage}

This specification of DPs follows the standard theoretical approach in a straightforward manner. However, it depends on two existential quantifiers that, throughout the proofs,  would bring several difficulties about which rule and position had created the DP being analyzed. This is because, due to the PVS proof calculus, whenever these existential quantifiers appear in the antecedent of a proof, their Skolemization leads to some arbitrary rule and position being chosen, making it difficult to construct derivations of terms associated with chained DPs.  It is easy to see that different \rhs\ positions, and even different rules can produce identical DPs; take for instance the TRS below, where $\langle h(x,y), g(x,y)\rangle$ can be built in three different manners.  
\[
\begin{array}{l@{\hspace{5mm}}l@{\hspace{5mm}}l}
\{h(x,y) \rightarrow h(g(x,y), g(g(x,y),y), &
h(x,y) \rightarrow g(x,y), &
g(x,y) \rightarrow y\}
\end{array}
\]

To discriminate the manner in which DPs are extracted from the rewrite rules and to circumvent the difficulties of existential quantifiers, an alternative notion of DP is provided in Specification \ref{spec:DPalt}. 

\hspace{-4mm}\begin{minipage}{.98\textwidth}\begin{footnotesize}
	\begin{lstlisting}[label = {spec:DPalt}, caption = {Predicate for Dependency Pairs as a pair of rule and position at its \rhs.}, mathescape, frame = single]
  dep_pair_alt?$(E)(e,\pi)$: bool =
    $e \in E \; \land \pi \in Pos(\rhs(e)) \; \land $ 
    app?$(\rhs(e)|_{\pi}) \;\land$ defined?$(E)($f$(\rhs(e)|_{\pi})$
	\end{lstlisting}
\end{footnotesize}
\end{minipage}

Having the rule and position that generate the DPs  allows, for instance, further specification of recursive functions  to adjust and accumulate the contexts of any infinite chain of DPs in order to build the associated infinite derivations (more details are given in Section \ref{sec:Necessity}). Here, it is important to stress that for termination analysis and automation, whenever {\tt dep\_pair\_alt?$(E)(e,\pi)$} and {\tt dep\_pair\_alt?$(E)(e',\pi')$} are such that $\lhs(e) = \;\lhs(e')$ and $\rhs(e)|_{\pi} = \;\rhs(e')|_{\pi'}$, it is sufficient to consider only one of these DPs. Other implementable refinements are discussed in Section \ref{sec:RelatedWork} on related work.  

In the remainder of the discussion, these two definitions will be distinguished if necessary, and in particular, for the sake of simplicity, the first and second elements of a DP will be identified with the \lhs\ of the rule and the subterm at position $\pi$ of the \rhs\ of the rule.

Notice that both specifications for DPs are currifyed, allowing the definition of the types {\tt  dep\_pair}$(E)$ and {\tt  dep\_pair\_alt}$(E)$.  

In order to check that an infinite sequence of DPs form an infinite (innermost) dependency chain, it is required, as given in Definitions \ref{def:DepChain} and \ref{def:inDepChain}, that every pair of consecutive DPs in this sequence be related through (innermost) non-root reductions, after renaming their variables, regarding some substitution. This gives rise to an imprecision since the type of substitutions does not allow infinite domains, as discussed in \cite{Sternagel10}. This issue is circumvented by specifying sequences DPs in association with sequences of substitutions. Thus, by allowing a different substitution for each DP in the sequence, it is possible to specify the notion of \emph{(innermost) chained DPs} (See Specification \ref{spec:DPinChain}).

\hspace{-4mm}\begin{minipage}{.98\textwidth}\begin{footnotesize}
	\begin{lstlisting}[label = {spec:DPinChain}, caption = {Predicates for (innermost) chainned Dependency Pairs.}, mathescape, frame = single]
  chained_dp?$(E)(dp_1,dp_2: $dep_pair$(E))(\sigma_1,\sigma_2)$: bool =
    $dp_1'2\sigma_1 \rightarrow^*_{>\lambda} dp_2'1\sigma_2$
    
  inn_chained_dp?$(E)(dp_1,dp_2: $dep_pair$(E))(\sigma_1,\sigma_2)$: bool =
    is_nr_normal_form?$(E)(dp_1'1\sigma_1) \;\land $ is_nr_normal_form?$(E)(dp_2'1\sigma_2) \; \land $
    $dp_1'2\sigma_1 \rightarrow^*_{in_{>\lambda}} dp_2'1\sigma_2$
  
	\end{lstlisting}
\end{footnotesize}
\end{minipage}

In Specification \ref{spec:DPinChain}, the elements of a DP, say $dp$, are projected by the operator $\_'\_$, as $dp'1$ and $dp'2$, used to project elements of tuples in PVS. Using these specifications of (innermost) chained DPs, whenever predicates in Specification \ref{spec:infInDPchains} hold for a pair of a sequence of DPs and substitutions, such pair is said to be an infinite (innermost) dependency chain.

\hspace{-4mm}\begin{minipage}{.98\textwidth}\begin{footnotesize}
	\begin{lstlisting}[label = {spec:infInDPchains}, caption={Predicates for infinite (innermost) Dependency Chains.},mathescape, frame = single]
  infinite_dep_chain?$(E)(dps: $sequence$[$dep_pair$(E)],$
                        $sigmas:$sequence$[$Sub$])$: bool = 
    $\forall(i:nat):$ chained_dp?$(E)(dps(i),dps(i+1))(sigmas(i),sigmas(i+1))$
      
  inn_infinite_dep_chain?$(E)(dps: $sequence$[$dep_pair$(E)],$
                            $sigmas:$sequence$[$Sub$])$: bool = 
    $\forall(i:nat):$ inn_chained_dp?$(E)(dps(i),dps(i+1))(sigmas(i),sigmas(i+1))$
  
	\end{lstlisting}
\end{footnotesize}
\end{minipage}

Finally, the (innermost) DP termination criterion is specified as the absence of such infinite chains in Specification \ref{spec:innDPterminations}, where the two first predicates specify the criterion for the standard notion of DPs (Specification \ref{spec:DPstd}), and the third and fourth ones for the alternative one (Specification \ref{spec:DPalt}). Notice that alternative DPs are translated into standard DPs in the third and fourth predicates. 

\hspace{-4mm}\begin{minipage}{.98\textwidth}\begin{footnotesize}
	\begin{lstlisting}[label = {spec:innDPterminations}, caption = {Predicates for (innermost) termination for the two specifications of DPs.}, mathescape, frame = single]
  dp_termination?$(E)$: bool = 
    $\forall (dps: $sequence$[$dep_pair$(E)], sigmas:$sequence$[Sub])$: 
      $\neg $infinite_dep_chain?$(E)(dps, sigmas) $
	
	
  inn_dp_termination?$(E)$: bool = 
    $\forall (dps: $sequence$[$dep_pair$(E)], sigmas:$sequence$[Sub])$: 
      $\neg $inn_infinite_dep_chain?$(E)(dps, sigmas) $
      
  dp_termination_alt?$(E)$: bool = 
    $\forall (dps\_alt: $sequence$[$dep_pair_alt$(E)], sigmas:$sequence$[Sub$]):
      LET $dps$ = LAMBDA($i$:nat): ($\lhs(dps\_alt(i)'1),$
                               $\rhs(dps\_alt(i)'1)|_{dps\_alt(i)'2}$) IN
      $\neg $infinite_dep_chain?$(E)(dps, sigmas) $
 
  inn_dp_termination_alt?$(E)$: bool = 
    $\forall (dps\_alt: $sequence$[$dep_pair_alt$(E)], sigmas:$sequence$[Sub$]):
      LET $dps$ = LAMBDA($i$:nat$): (\lhs(dps\_alt(i)'1),$
                              $ rhs(dps\_alt(i)'1)|_{dps\_alt(i)'2}$) IN
      $\neg $inn_infinite_dep_chain?$(E)(dps, sigmas) $
	
	\end{lstlisting}
\end{footnotesize}
\end{minipage}

As aforementioned, several elements were specified to deal with various reduction relations, for which several properties were formalized  but not discussed in this paper since the focus here is on formalization of innermost termination by DPs. Furthermore, the alternative version of DPs is used aiming to simplify proofs, and in order to ensure that the corresponding innermost DP criteria are the same, the equivalence {\tt inn\_dp\_termination?(E)} $\Leftrightarrow$ {\tt inn\_dp\_termination\_alt?(E)}  was formalized in theory {\tt dependency\_pairs.pvs} as lemma {\tt dp\_termination\_and\_alt\_eq}. This proof is quite simple, building by contraposition infinite sequences of standard chained DPs from alternative ones and vice versa.

\section{Necessity for the Innermost Dependency Pairs Criterion}\label{sec:Necessity}

Lemma {\tt inn\_noetherian\_implies\_inn\_dp\_termination} formalizes this result, which is specified in Specification \ref{spec:noetherianANDnecessityLemma} along with the specification of the {\tt noetherian?} predicate over a given relation, which specified as holding whenever the converse of this relation is well-founded (both {\tt well\_founded?} predicate and function {\tt converse} follow the standard definition and are specified in the prelude file of PVS). 

\hspace{-4mm}\begin{minipage}{.98\textwidth}\begin{footnotesize}
	\begin{lstlisting}[label = {spec:noetherianANDnecessityLemma}, caption = {The {\tt noetherian?} predicate and the necessity lemma.}, mathescape, frame = single]
  noetherian?(R): bool = well_founded?(converse(R))
  
  inn_noetherian_implies_inn_dp_termination: LEMMA 
    $\forall (E)$: 
      noetherian?(innermost_reduction?$(E)) \rightarrow$ inn_dp_termination?$(E)$
	\end{lstlisting}
\end{footnotesize}
\end{minipage}

\begin{figure}[H]

\begin{tikzpicture}[scale=0.72, transform shape]

\begin{scriptsize}

\draw (-1.5,1) -- (-2,0) -- (-1.5,-1); 
\node[align=center] at (-0.9,0.8) {$l_1\sigma_1$};
\nSPolygon{3}{1.5}{olive}{olive!20}{-0.9,-0.15}{}
\node[align=center] at (0.2,0) { $\overset{\lambda}{\longrightarrow}_{in}$};
\node[align=center] at (1.3,0.8) {$r_1\sigma_1$};
\nSPolygon{3}{1.5}{black}{white}{1.3,-0.15}{}
\node[align=center] at (1.35,-1) {$r_1\sigma_1|_{\pi_1}$ };
\nSPolygon{3}{0.7}{black}{olive!20}{1.3,-0.5}{};
\draw [mbrace = 4] (0.9,-0.7) -- (1.7,-0.7);
\draw (2,1) -- (2.5,0) -- (2,-1); 

\node[align=center] at (2.9,-0.5) { $\longrightarrow^*_{in_{>\lambda}}$};

\draw (3.7,1) -- (3.2,0) -- (3.7,-1); 
\node[align=center] at (4.3,0.8) {$l_2\sigma_2$};
\nSPolygon{3}{1.5}{olive}{olive!20}{4.3,-0.15}{}
\nHPolygon{3}{1.5}{olive}{north east lines}{black}{4.3,-0.15}{}
\node[align=center] at (5.4,0) { $\overset{\lambda}{\longrightarrow}_{in}$};
\node[align=center] at (6.5,0.8) {$r_2\sigma_2$};
\nSPolygon{3}{1.5}{black}{gray!20}{6.5,-0.15}{}
\node[align=center] at (6.5,-1) {$r_2\sigma_2|_{\pi_2}$ };
\nSPolygon{3}{0.7}{black}{olive!20}{6.5,-0.5}{};
\nHPolygon{3}{0.7}{black}{north east lines}{black}{6.5,-0.5}{};
\draw [mbrace = 4] (6.1,-0.7) -- (6.9,-0.7);

\draw (7.2,1) -- (7.7,0) -- (7.2,-1); 

\node[align=center] at (8.2,-0.5) { $\longrightarrow^*_{in_{>\lambda}}$};

\draw (9,1) -- (8.5,0) -- (9,-1); 
\node[align=center] at (9.6,0.8) {$l_3\sigma_3$};
\nSPolygon{3}{1.5}{olive}{olive!20}{9.6,-0.15}{}
\nHPolygon{3}{1.5}{olive}{vertical lines}{black}{9.6,-0.15}{}
\node[align=center] at (10.6,0) { $\overset{\lambda}{\longrightarrow}_{in}$};
\node[align=center] at (11.7,0.8) {$r_3\sigma_3$};
\nSPolygon{3}{1.5}{black}{blue!30}{11.7,-0.15}{}
\node[align=center] at (11.7,-1) {$r_3\sigma_3|_{\pi_3}$ };
\nSPolygon{3}{0.7}{black}{olive!20}{11.7,-0.5}{};
\nHPolygon{3}{0.7}{black}{vertical lines}{black}{11.7,-0.5}{};
\draw [mbrace = 4] (11.3,-0.7) -- (12.1,-0.7);

\draw (12.4,1) -- (12.9,0) -- (12.4,-1); 

\node[align=center] at (14,-0.5) { $\longrightarrow^*_{in_{>\lambda}} \; \cdots$};

\draw [mbrace = 4] (-2,-1.5) -- (15,-1.5);
\node[align=center] at (6.5,-2) { $\Downarrow$};


\node[align=center] at (3,-2.5) {$r_1\sigma_1$};
\nSPolygon{3}{1.5}{black}{white}{3,-3.5}{}
\nSPolygon{3}{0.7}{black}{olive!20}{3,-3.9}{};

\node[align=center] at (4.5,-3.5) { $\longrightarrow^+_{in}$};

\node[align=center] at (6,-2.5) {$r_1\sigma_1[\pi_1 \leftarrow r_2\sigma_2]$};
\nSPolygon{3}{1.5}{black}{white}{6,-3.5}{}
\nSPolygon{3}{1.2}{black}{gray!20}{6,-4.1}{};
\nSPolygon{3}{0.7}{olive}{olive!20}{6,-4.5}{};
\nHPolygon{3}{0.7}{black}{north east lines}{black}{6,-4.5}{};

\node[align=center] at (8,-3.5) { $\longrightarrow^+_{in}$};
\node[align=center] at (10,-2.5) {$r_1\sigma_1[\pi_1 \leftarrow r_2\sigma_2[\pi_2 \leftarrow r_3\sigma_3]]$};
\nSPolygon{3}{1.5}{black}{white}{10,-3.5}{}
\nSPolygon{3}{1.2}{black}{gray!20}{10,-4.1}{};
\nSPolygon{3}{1}{black}{blue!30}{10,-4.6}{}
\nSPolygon{3}{0.7}{olive}{olive!20}{10,-4.9}{};
\nHPolygon{3}{0.7}{black}{vertical lines}{black}{10,-4.9}{};

\node[align=center] at (12,-3.5) { $\longrightarrow^+_{in} \; \cdots$};


\end{scriptsize}
\end{tikzpicture}
	\caption{Proof sketch: building infinite innermost derivations from infinite innermost DP-chains.}\label{fig:infDPinfDeriv}
\end{figure}
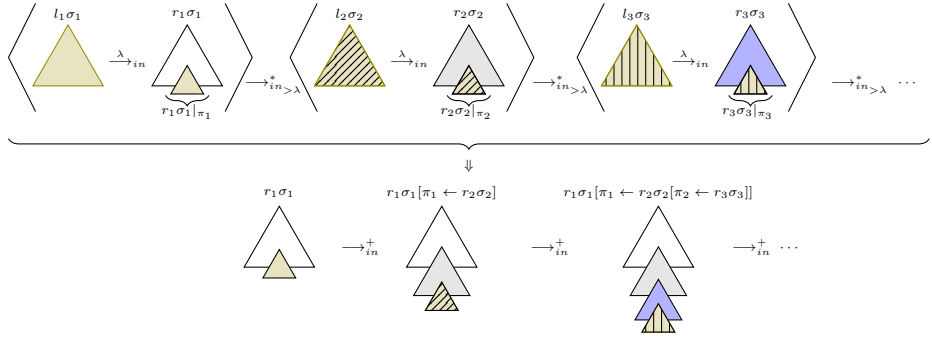

The formalization follows by contraposition, by building an infinite sequence of terms associated with an infinite innermost derivation from an infinite chain of dependency pairs. In order to build these terms, it is necessary to accumulate the contexts where the reductions would take place regarding the \rhs\ of the rule that generates each DP in the chain. The intuition of this formalization follows directly from the theory,  and  is summarized in the sketch given in Figure \ref{fig:infDPinfDeriv}.

 Since there is a root reduction associated with each DP in the sequence, from its \lhs\ to the  \rhs\ of the related rule, and a non-root innermost derivation  to reach the \lhs\ of the next DP from the \rhs\ of the current DP, it is relatively simple to manipulate the rules and positions using the  alternative dependency chain specification to  build recursively a sequence of terms related by $\rightarrow^+_{i}$ through the replacement operation.  

To perform this construction, the recursive function {\tt term\_pos\_dps\_alt} is used, taking sequences of DPs and substitutions and producing indexed pairs of term and position accumulating contexts in such a way that the terms are related by $\rightarrow^+_{i}$ whenever the given sequence is chained (Specification \ref{spec:DPtoTermPos}).  As illustrated in Figure~\ref{fig:infDPinfDeriv}, if the sequence is chained,  the first pair of term and position is computed as $(r_1\sigma_1, \pi_1)$;  the second as $(r_1[\pi_1\leftarrow r_2\sigma_2], \pi_1\circ\pi_2)$; and so on.   The function {\tt term\_pos\_dps\_alt} uses the previously obtained accumulated context ($C$) and replaces the \rhs\ of the current DP by the \rhs\ of the next DP in the sequence. Positions to perform the replacement are given by accumulation of the positions in the alternative definition of DPs ($\pi$).

\hspace{-4mm}\begin{minipage}{.98\textwidth}\begin{footnotesize}
	\begin{lstlisting}[label = {spec:DPtoTermPos}, caption = {Function to accumulate contexts to build an infinite sequence of terms.}, mathescape, frame = single]
  term_pos_dps_alt$(E)(dps: $sequence$[$dep_pair_alt$(E)],$ 
	             $sigmas:$sequence$[$Sub$], i:$nat): 
	             RECURSIVE {$(C,\pi)\; |\; \pi \in Pos(C)$}=
    IF $i=0$ THEN 
      ($\rhs(dps(0)'1)sigmas(i),dps(0)'2$)
    ELSE LET $(C,\pi)$ = term_pos_dps_alt$(E)(dps,sigmas,i-1)$ IN 
      $(C[\pi \leftarrow \rhs(dps(i)'1)sigmas(i)], \pi \circ dps(i)'2)$
    ENDIF
  MEASURE $i$
	\end{lstlisting}
\end{footnotesize}
\end{minipage}

Then, an infinite sequence of terms can be built from an infinite chain given by sequences  of DPs and substitutions $dps$ and  $sigmas$ as: 

\hspace{-4mm}\begin{minipage}{.98\textwidth}\begin{footnotesize}
	\begin{lstlisting}[label = {spec:infTerms}, caption = {Function to build the terms in an infinite derivation.}, mathescape, frame = single]
  LAMBDA($i$:nat):  term_pos_dps_alt$(E)(dps,sigmas,i)'1$
	\end{lstlisting}
\end{footnotesize}
\end{minipage}

 Notice that the function {\tt term\_pos\_dps\_alt} would provide an infinite sequence of terms for any pair of infinite sequences of DPs and substitutions,  disregarding if they  form an infinite innermost chain or not. 
 To prove that the generated infinite sequence  indeed describes an infinite derivation for the relation  $\rightarrow_{i}$, this function should be applied to a pair $dps$ and $sigmas$ that constitutes an infinite chain.

This is proved by showing the non-noetherianity of $\rightarrow^+_{i}$ that relates consecutive terms generated by the function {\tt term\_pos\_dps\_alt}. The proof follows by induction, whereas for the induction basis it must be proved that the first term generated is related to the second by $\rightarrow^+_{i}$. {\tt term\_pos\_dps\_alt} builds these terms just using the first and second DPs and substitutions, say $((l_1, r_1), \pi_1)$, $((l_2, r_2), \pi_2)$,  and $\sigma_1$ and $\sigma_2$ as in Figure \ref{fig:infDPinfDeriv}, in the chained input. The first term is $r_1\sigma_1$ and the second $r_1\sigma_1[\pi_1 \leftarrow r_2\sigma_2]$, which is equal  to  $r_1\sigma_1[\pi_1 \leftarrow l_2\sigma_2[\lambda \leftarrow r_2\sigma_2]]$. Since contiguous pairs in the sequence are innermost chained and $\rightarrow^*_{in_{>\lambda}}$ is compatible with contexts (by monotony of closures, since $\rightarrow_{i}$ is compatible with contexts and $\overset{>\lambda}{\rightarrow}_{i} \subseteq \rightarrow_{i}$), one has that $r_1\sigma_1 \rightarrow^*_{in_{>\lambda}} r_1\sigma_1[\pi_1 \leftarrow l_2\sigma_2]$. And, also by the innermost chained property, $l_2\sigma_2$ is a normal instance of the \lhs\ of a rule, i.e., a single innermost reduction step can be applied only at root position giving $r_2\sigma_2$. Since a single innermost reduction step corresponds directly to a replacement operation, and in this case at root position, one would have one innermost reduction step $r_1\sigma_1[\pi_1 \leftarrow l_2\sigma_2] \overset{\pi_1}{\rightarrow}_{i} r_1\sigma_1[\pi_1 \leftarrow r_2\sigma_2]$. Thus, one would have $r_1\sigma_1 \rightarrow^+_{i} r_1\sigma_1[\pi_1 \leftarrow r_2\sigma_2]$.  The inductive step considers analogously contiguous DPs and substitutions in the chained input, the only extra details are regarding the current term and position computed in the previous recursive step by  {\tt term\_pos\_dps\_alt}.  Notice that in the $i^{th}$ iteration the current term can be seen as a context $C$ with a hole at the accumulated position, say $\pi$, filled with term $r_i|_{\pi_i}\sigma_i$. Indeed, in the induction basis the context is given by $r_1\sigma_1$ with a hole at position $\pi_1$.  The term and accumulated position generated by {\tt term\_pos\_dps\_alt} are given as $C[\pi\leftarrow  r_{i+1}\sigma_{i+1}]$ and $\pi\circ\pi_{i+1}$. Notice that this term can be seen as a context with a hole at the accumulated position filled with the term $r_{i+1}|_{\pi_{i+1}}$. Finally, observe that  $C[r_i|_{\pi_i}\sigma_i]  \rightarrow^+_{i} C[r_{i+1}\sigma_{i+1}]$.

Notice that this formalization is very similar to its pen-and-paper version, disregarding the specification. However, the construction of an actual function to generate each pair of accumulated context and position simplifies the inductive and constructive proof of  the existence of the infinite derivation. Furthermore, proof elements that can seem too trivial must be precisely used, such as the mentioned closure of context, monotony of closures, subset properties and properties regarding composition of positions in replacements. For example, the last property is used in proving correctness of the {\it predicate subtyping} condition   $\{(C,\pi) \;|\; \pi \in \mbox{\it Pos}(C) \}$ of the pairs built by the function {\tt term\_pos\_dps\_alt} (these aspects are  discussed in detail in Section \ref{ssec:BuildingDPsSubs}). These properties are formalized in the PVS theory {\tt TRS} in a general manner allowing its application for arbitrary rewriting relations.

\section{Sufficiency for the Innermost Dependency Pairs Criterion}\label{sec:Sufficiency}

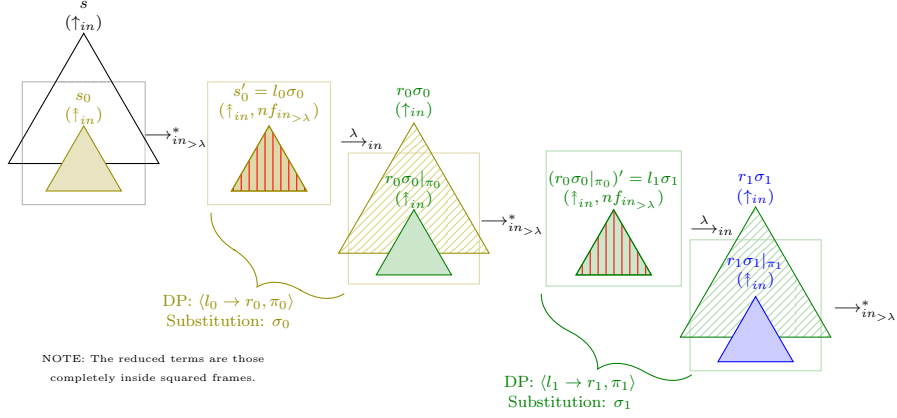
\begin{figure}[H]
	\begin{tikzpicture}[scale=0.77, transform shape]
\begin{footnotesize}
\node[align=center] at (-0.7,-1.9) {$s$ \\ $(\uparrow_{in})$};
\nSPolygon{4}{3}{black!30}{white}{-0.7,-4.1}{}
\nHPolygon{3}{3}{black}{vertical lines}{white}{-0.7,-3.7}{};
{\color{black!30}\draw  (-1.15,-3.04)--(-0.25,-3.04);}
\node[align=center] at (-0.7,-3.5) {\color{olive}$s_0$ \\ \color{olive}$(\mnt_{in})$};
\nSPolygon{3}{1.5}{olive}{olive!20}{-0.7,-4.55}{};

\node at (0.9,-4) {$\longrightarrow^*_{in_{> \lambda}}$};

\nSPolygon{4}{3}{olive!30}{white}{2.5,-4.1}{}
\node[align=center] at (2.5,-3.4) {\color{olive}$s_0' = l_0 \sigma_0$ \\ \color{olive}$(\mnt_{in}, nf_{in_{>\lambda}})$};
\nSPolygon{3}{1.5}{olive}{olive!30}{2.5,-4.55}{};
\nHPolygon{3}{1.5}{olive}{vertical lines}{red}{2.5,-4.55}{};

\node at (4.1,-4) {$\overset{ \lambda}{\longrightarrow}_{in}$};

\node[align=center] at (5,-3.4) {\color{Green}$ r_0\sigma_0$ \\ \color{Green}$(\uparrow_{in})$};
\nSPolygon{4}{3.2}{olive!30}{white}{5,-5.4}{}
\nHPolygon{3}{3}{olive}{north east lines}{olive!30}{5,-5.25}{};
\node[align=center] at (5,-4.9) {\color{Green}$r_0\sigma_0 |_{\pi_0}$ \\ \color{Green}$(\mnt_{in})$};
\nSPolygon{3}{1.5}{Green}{Green!20}{5,-6}{};

{\color{olive}\draw[mbrace = 15] (1.5,-5.3) -- (3.8,-6.7);}
\node[align=center] at (1.8,-7) {\color{olive}DP: $\langle l_0 \rightarrow r_0, \pi_0\rangle$ \\ \color{olive}Substitution: $\sigma_0$};

\node at (6.7,-5.5) {$\longrightarrow^*_{in_{> \lambda}}$};

\nSPolygon{4}{3.3}{Green!30}{white}{8.45,-5.4}{}
\node[align=center] at (8.45,-4.9) {\color{Green}$(r_0\sigma_0|_{\pi_0})' = l_1 \sigma_1$ \\ \color{Green}$(\mnt_{in}, nf_{in_{>\lambda}})$};
\nSPolygon{3}{1.5}{Green}{Green!20}{8.45,-6}{};
\nHPolygon{3}{1.5}{Green}{vertical lines}{red}{8.45,-6}{};

\node at (10.15,-5.5) {$\overset{ \lambda}{\longrightarrow}_{in}$};


\node[align=center] at (10.9,-4.9) {\color{blue}$r_1\sigma_1$ \\ \color{blue}$(\uparrow_{in})$};
\nSPolygon{4}{3.2}{Green!30}{white}{10.9,-6.9}{}
\nHPolygon{3}{3}{Green}{north east lines}{Green!30}{10.9,-6.7}{};
\node[align=center] at (10.9,-6.3) {\color{blue}$r_1\sigma_1|_{\pi_1}$ \\ \color{blue}$(\mnt_{in})$};
\nSPolygon{3}{1.5}{blue}{blue!20}{10.9,-7.5}{};

{\color{Green}\draw[mbrace = 15] (7.2,-6.7) -- (9.8,-8.1);}
\node[align=center] at (7.7,-8.4) {\color{Green}DP: $\langle l_1 \rightarrow r_1, \pi_1\rangle$ \\ \color{Green} Substitution: $\sigma_1$};

\node at (12.8,-7) {$\longrightarrow^*_{in_{> \lambda}}$};

\node[align=center] at (0.5,-8) {\tiny NOTE: The reduced terms are those \\ \tiny completely inside squared frames.};

\end{footnotesize}
\end{tikzpicture}
	\caption{Proof sketch: building infinite innermost DP-chains from infinite innermost derivations. Notice that the two DPs created, along with their respective substitutions, form chained DPs.}\label{fig:infDerivInfDP}
\end{figure}

The formalization is by contraposition. The core of the proof follows the idea  in \cite{ArtsGiesl00} to construct infinite chains from infinite innermost derivations. In an implementional level, to go from infinite derivations to infinite sequences of DPs that would create an infinite chain is challenging. Indeed, constructing the DPs requires, initially, choosing \emph{mint} subterms from those terms leading to infinite innermost derivations; afterwards, choosing non-root innermost normalized terms;  and, finally,  choosing instances of rules that apply at root positions of these terms from which DPs can be constructed. All these choices are based on existential proof techniques.  Figure \ref{fig:infDerivInfDP} illustrates the main steps of the kernel of the construction of chained DPs:

\begin{itemize}
	\item Existence of \emph{mint} subterms of innermost non-terminating terms is represented as the small triangles inside big ones.  This part of the development is explained in Subsection \ref{ssec:ExistsMint}.
	\item Existence of non-root innermost normalized terms obtained by  derivations (through relation $\overset{>\lambda}{\rightarrow}_{i}$) from these \emph{mint} subterms, represented as vertically striped triangles, is detailed in Subsection \ref{ssec:InNormalFromMint}.
	\item Existence of DPs from rules and substitutions that reduce non-root innermost normalized terms at root position, which also are innermost non-terminating, into innermost non-terminating terms.  The DPs are represented by pairs of small vertically striped and small plain triangles and the latter by reductions (through relation $\overset{\lambda}{\rightarrow}_{i}$) from vertically to diagonally striped triangles. This result is explained in Subsection \ref{ssec:ExistenceOfDP}.
	
\end{itemize}

The last step of the construction illustrated in Figure \ref{fig:infDerivInfDP} permits, as the first one, application of a lemma of existence of \emph{mint} subterms (for innermost non-terminating terms). In the last step, this result will allow constructing the required DPs.

Subsection \ref{ssec:BuildingDPsSubs} then discusses how getting adequate pairs of consecutive chained DPs and associated normal substitutions, and Subsection \ref{ssec:BuldingInfInDepChain}, finally,  details the construction of the required chain of DPs.  

\subsection{Existence of \emph{mint} Subterms}\label{ssec:ExistsMint}

The \emph{mint} property ($\mnt_{i}$) over terms is provided in the Specification \ref{spec:mintANDexistsmint} by predicate {\tt minimal\_non\_innermost\_terminating?}. Also in this box one has the specification of lemma {\tt  inn\_non\_terminating\_has\_mint}, whose formalization ensures the existence of \emph{mint} subterms regarding innermost non-terminating terms. The proof follows by induction on the structure of the term. The induction basis is trivial since variable terms are not reducible, so variables cannot give rise to infinite derivations. For the inductive step, whenever the term $t$ has an empty list of arguments (that is, $t$ is a constant), the only position it has is its root, thus, the \emph{mint} subterm is the term itself; otherwise, either all its proper subterms are innermost terminating and then the  term itself is  \emph{mint} or, by induction hypothesis, some of its arguments is innermost non-terminating, say its $i{th}$ argument, and then it has a \emph{mint} subterm at some position $\pi$,  thus, the \emph{mint} subterm of $t$ is chosen as $t|_{i \pi}$.  

\hspace{-4mm}\begin{minipage}{.98\textwidth}\begin{footnotesize}
	\begin{lstlisting}[label = {spec:mintANDexistsmint}, caption = {Predicate for specifying \emph{mint} terms and lemma over existence of \emph{mint} subterms in innermost non-terminating terms.}, mathescape, frame = single]	
  minimal_innermost_non_terminating?(E)(t:term): bool = 
    $\uparrow_{i}$(t) $\land$ $\forall$($\pi \in Pos(t) | \pi \neq \lambda$): SN$_{i}(t|_{\pi})$)
  
  inn_non_terminating_has_mint: LEMMA
    $\forall$(E)(t: term | $\uparrow_{i}$(t)):  $\exists$($\pi \in Pos(t))):  \;\mnt_{i}(t|_{\pi}$))
	
	\end{lstlisting}
\end{footnotesize}
\end{minipage}

\subsection{Non-root Innermost Normalization of \emph{mint} Terms}\label{ssec:InNormalFromMint}

The second step in the formalization proves that every \emph{mint} term can be non-root innermost normalized (into an innermost non-terminating term). This result appears to be, as given in analytic proofs, a simple observation. By definition, every proper subterm of a \emph{mint} term is innermost terminating, and consequently  no argument of this term may give rise to an infinite innermost derivation. However, formalizing such result by contradiction requires several auxiliary functions and lemmas related to structural properties of such derivations that also consider positions and arguments in which each reduction step happens.  These technicalities of the formalization are necessary to obtain a key result that assuming the existence of an infinite non-root innermost derivation from a \emph{mint} term guarantees that some of its arguments begin an infinite innermost derivation, which gives the contradiction.  

\subsubsection{\emph{mint} Terms are Non-root Innermost Terminating}

For the remainder of this subsection, consider elements on Specification \ref{spec:elements}, where $s$, $seqt$ and $seqp$  are fixed term, sequences of terms and positions, respectively, associated with an infinite non-root innermost derivation on non-root innermost descendants of $s$, such that the $n^{th}$ term in the sequence $seqt$ reduces into the $(n+1)^{th}$ term at position $seqp(n)$.  Also, $l$ will denote a valid argument of $s$ (and as it will be seen, also a valid argument of any of its descendants).

\hspace{-4mm}\begin{minipage}{.98\textwidth}\begin{footnotesize}
	\begin{lstlisting}[label = {spec:elements}, caption = {Fixed term, argument position of the term, and sequences of terms and positions used in the formalization.}, mathescape, frame = single]
$s$:term | app?(s)
	
$l$ : posnat |  $l \leq$ length(args($s$))
	
$seqt$: sequence[term] | $\forall(n$: nat): $s \overset{>\lambda}{\rightarrow}_{i} seqt(n)$
	
$seqp$: sequence[position] | $\forall(n: $nat$): 
	                            seqp(n) \in Pos(seqt(n)) \;\land $
	                           $seqp(n)\neq \lambda \;\land seqt(n)\overset{seqp(n)}{\rightarrow_{i}}seqt(n+1)$
	
\end{lstlisting}
\end{footnotesize}
\end{minipage}

The predicate {\tt inf\_red\_arg\_in\_inf\_nr\_im\_red} in Specification \ref{spec:infReductionArg} holds whenever for a sequence of positions there is an infinite number of positions in the sequence starting with the same natural.  For $seqp$ and $l$ as in Specification \ref{spec:elements},  this predicate will be applied to state the existence of an infinite set of indices in the sequence of terms $seqt$ in which the reduction happens at the  $l^{th}$ argument.   The function {\tt args\_of\_pos\_seq} is just used to give the argument of each position in a sequence of positions.
 
\hspace{-4mm}\begin{minipage}{.98\textwidth}\begin{footnotesize}
	\begin{lstlisting}[label = {spec:infReductionArg}, caption = {Function to extract the argument position from a given position in a sequence of positions where reductions take place and predicate for checking if there exist infinite reductions at a given argument position.}, mathescape, frame = single]
args_of_pos_seq($seq$: sequence[position] | $\forall(i:$nat$): seqp(i) \neq \lambda)$
               $(n:nat) : $ posnat = first($seqp(n)$)	
	
inf_red_arg_in_inf_nr_im_red($seq$: sequence[position] |
                                  $\forall(i:$nat$):seqp(i) \neq \lambda)$
                            $(i : $posnat): bool =
      is_infinite(inverse_image(args_of_pos_seq($seq$), $i$))
	
	\end{lstlisting}
\end{footnotesize}
\end{minipage}

Then, for any $l$-th argument of the given term $s$ such that the predicate {\tt inf\_red\_arg\_in\_inf\_nr\_im\_red$(seqt)(l)$} holds, the  function {\tt nth\_index} (Specification \ref{spec:nthindex}) provides the index of the sequence in which the  $(n+1)^{th}$ reduction at argument $l$ happens.  
 
 \hspace{-4mm}\begin{minipage}{.98\textwidth}\begin{footnotesize} 	
\begin{lstlisting}[label = {spec:nthindex}, caption = {Function {\tt nth\_index}.}, mathescape, frame = single]
nth_index$(E)(s)(seqt)(seqp)(l)(n:nat)$ : nat = 
  choose({$m$: nat | args_of_pos_seq$(seqp)(m) = l \;\land$
                    card({$k$: nat | args_of_pos_seq$(seqp)(k) = l \land $ 
                                   $k < m $}) $= n$}) 
\end{lstlisting}
 \end{footnotesize}
\end{minipage}

Notice that well-definedness of these functions is a consequence of the type of $l$ that is a dependent type satisfying the predicate  {\tt inf\_red\_arg\_in\_inf\_nr\_im\_red}, which means that reductions at the $l^{th}$ argument happen infinitely many times. The main technical difficulty of formalizing well-definedness is related to  guaranteeing non-emptiness of the argument of the built-in function {\tt choose}.   This constraint is fulfilled by the auxiliary lemma {\tt exists\_nth\_in\_inf\_nr\_im\_red} in Specification \ref{spec:nonEmptyArgument}. 

\hspace{-4mm}\begin{minipage}{.98\textwidth}\begin{footnotesize} 	
	\begin{lstlisting}[label = {spec:nonEmptyArgument}, caption = {Non-emptiness lemma for the argument positions where infinite reductions may take place.}, mathescape, frame = single]
  exists_nth_in_inf_nr_im_red : LEMMA
    $\forall (n : $nat$) :  \exists (m : $nat$) :$
      args_of_pos_seq$(seqp)(m) = l \;\land$
      card$(\{k: $nat | args_of_pos_seq$(seqp)(k) = l \land k < m \}) = n $
	\end{lstlisting}
\end{footnotesize}
\end{minipage}

The formalization of this lemma follows by induction on $n$ and, although simple, requires several auxiliary lemmas over sets. In the induction basis, since one has infinite reductions at argument $l$, the set of indices where such reductions take place is infinite, and thus, nonempty (by application of the PVS prelude lemma {\tt infinite\_nonempty}). Thus, it is possible to use PVS function {\tt min} (over nonempty sets) to choose the smallest index of this set. By the definition of this {\tt min} function, it is ensured that the set of indices  smaller than this minimum in this set is empty, and thus has cardinality zero (by applying PVS prelude lemma {\tt card\_empty?}). For the inductive step, one must provide the index where one has a reduction at argument $l$ such that it has exactly $n+1$ indices smaller than it where reductions at argument $l$ occur. By induction hypothesis, there exists an index $m$ for which reduction take place at argument $l$, and for which the cardinality of indices smaller than $m$ with reductions at argument $l$ is $n$.  Thus, the required index is built as the minimum index bigger than $m$ for which the reduction happens at argument $l$. Correctness  of such indices follows similarly to the induction basis. First, since the predicate  {\tt inf\_red\_arg\_in\_inf\_nr\_im\_red} holds, it is possible to ensure that the set of indices greater than index $m$ for which reductions happen at argument $l^{th}$ is infinite, which allows application of the function {\tt min}. Then one builds an equivalent set to the one of all indices smaller than this minimum as the addition of index $m$ to the set of indices smaller than  $m$ (where one has reductions at argument $l^{th}$). This construction allows one to use another prelude lemma regarding cardinality of addition of elements in finite sets ({\tt card\_add}) to state that the cardinality of this new set is  $n + 1$. 

 Soundness of {\tt nth\_index} follows from auxiliary properties such as its monoto- ny and \emph{completeness}, the latter meaning that this function covers exactly (all) the  indices in which reductions happen at the $l^{th}$ argument. The formalization of these properties follows directly from the conditions fulfilled by the natural numbers chosen as the indices in  {\tt nth\_index} and prelude lemmas over cardinality of subsets ({\tt card\_subset}), since each index provided gives rise to a subset of the next one. These properties allow an easy formalization of a useful auxiliary result stating that for every index of $seqt$ below {\tt nth\_index($0$)} and between {\tt nth\_index($i$)}$+1$  and {\tt nth\_index($i+1$)}  there are no reductions in the $l^{th}$ argument (lemma {\tt argument\_protected\_in\_non\_nth\_index}). And then it is possible to ensure that there are only finitely many non-root innermost reductions regarding a term with \emph{mint} property, which is stated in Specification \ref{spec:mintNRIterm} as the lemma {\tt mint\_is\_nr\_inn\_terminating}.

\hspace{-4mm}\begin{minipage}{.98\textwidth}\begin{footnotesize}
\begin{lstlisting}[label = {spec:mintNRIterm}, caption = {Lemma for non-root innermost termination of \emph{mint} terms.}, mathescape, frame = single]
  mint_is_nr_inn_terminating: LEMMA  $\mnt_{i}(s) \rightarrow  $noetherian?$(\underset{s}{\rightarrow} \,_{in_{>\lambda}}))$
\end{lstlisting}
\end{footnotesize}
\end{minipage}

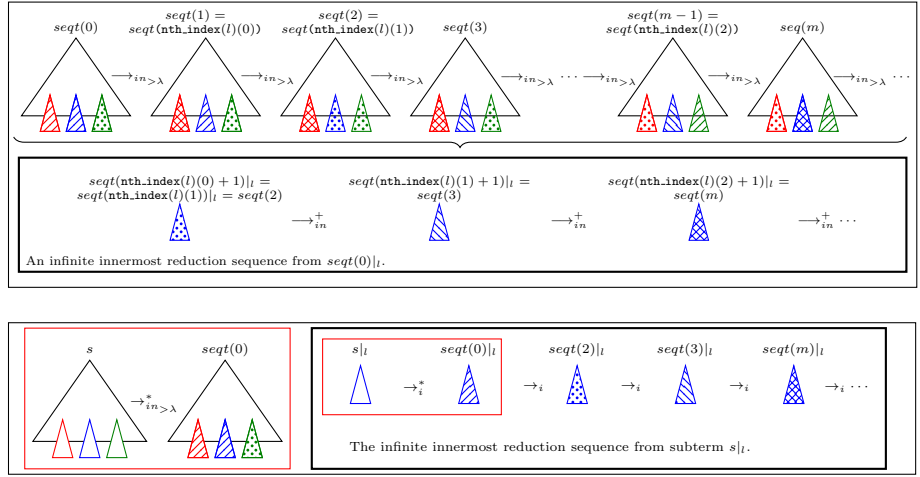
\begin{figure}[H]


\begin{tikzpicture}[scale=0.69, transform shape]

\begin{scriptsize}


\draw (-1.3,.5) rectangle (16.2,-5); 

\node[align=center] at (0,0) {$seqt(0)$};
\drawterm{black}{1.5}{70}{white}{0,-1.7}
\drawterm{red}{.7}{30}{white}{-.5,-2}
\drawHterm{red}{.7}{30}{north east lines}{red}{-.5,-2}
\drawterm{blue}{.7}{30}{white}{0,-2}
\drawHterm{blue}{.7}{30}{north east lines}{blue}{0,-2}
\drawterm{Green}{.7}{30}{white}{0.5,-2}
\drawHterm{Green}{.7}{30}{crosshatch dots}{Green}{0.5,-2}

\node at (1.2,-1) {$\longrightarrow_{in_{> \lambda}}$};

\node[align=center] at (2.3,0.1) {$seqt(1) = $\\ {\tt $seqt$(nth\_index$(l)(0)$)}};
\drawterm{black}{1.5}{70}{white}{2.5,-1.7}
\drawterm{red}{.7}{30}{white}{2,-2}
\drawHterm{red}{.7}{30}{crosshatch}{red}{2,-2}
\drawHterm{blue}{.7}{30}{north east lines}{blue}{2.5,-2}
\drawterm{Green}{.7}{30}{white}{3,-2}
\drawHterm{Green}{.7}{30}{crosshatch dots}{Green}{3,-2}
\node at (3.7,-1) {$\longrightarrow_{in_{> \lambda}}$};

\node[align=center] at (5.25,0.1) {$seqt(2) = $\\ {\tt $seqt$(nth\_index$(l)(1)$)}};
\drawterm{black}{1.5}{70}{white}{5,-1.7}
\drawterm{red}{.7}{30}{white}{4.5,-2}
\drawHterm{red}{.7}{30}{crosshatch}{red}{4.5,-2}
\drawHterm{blue}{.7}{30}{crosshatch dots}{blue}{5,-2}
\drawHterm{Green}{.7}{30}{crosshatch dots}{Green}{5.5,-2}
\node at (6.2,-1) {$\longrightarrow_{in_{> \lambda}}$};

\node[align=center] at (7.5,0) {$seqt(3)$};
\drawterm{black}{1.5}{70}{white}{7.5,-1.7}
\drawterm{red}{.7}{30}{white}{7,-2}
\drawHterm{red}{.7}{30}{crosshatch}{red}{7,-2}
\drawHterm{blue}{.7}{30}{north west lines}{blue}{7.5,-2}
\drawHterm{Green}{.7}{30}{crosshatch dots}{Green}{8,-2}

\node at (9.5,-1) {$\longrightarrow_{in_{> \lambda}} \cdots \longrightarrow_{in_{> \lambda}}$};

\node[align=center] at (11.5,0.1) {$seqt(m-1)=$\\ {\tt $seqt$(nth\_index$(l)(2)$)}};
\drawterm{black}{1.5}{70}{white}{11.5,-1.7}
\drawterm{red}{.7}{30}{white}{11,-2}
\drawHterm{red}{.7}{30}{crosshatch dots}{red}{11,-2}
\drawHterm{blue}{.7}{30}{north west lines}{blue}{11.5,-2}
\drawHterm{Green}{.7}{30}{north east lines}{Green}{12,-2}

\node at (12.7,-1) {$\longrightarrow_{in_{> \lambda}}$};

\node[align=center] at (14,0) {$seq(m)$};
\drawterm{black}{1.5}{70}{white}{14,-1.7}
\drawterm{red}{.7}{30}{white}{13.5,-2}
\drawHterm{red}{.7}{30}{crosshatch dots}{red}{13.5,-2}
\drawterm{blue}{.7}{30}{white}{14,-2}
\drawHterm{blue}{.7}{30}{crosshatch}{blue}{14,-2}
\drawHterm{Green}{.7}{30}{north east lines}{Green}{14.5,-2}
\node at (15.35,-1) {$\longrightarrow_{in_{> \lambda}} \cdots$};

\draw [mbrace = 4] (-1.2,-2.1) -- (16,-2.1);

\draw[thick] (-1.1,-2.5) rectangle (16,-4.7); 

\node[align=center] at (2,-3.1) {{\tt $seqt($nth\_index$(l)(0) + 1)|_{l} = $} \\ {\tt $seqt($nth\_index$(l)(1))|_{l} = seqt(2)$ }};
\drawHterm{blue}{.7}{30}{crosshatch dots}{blue}{2,-4.1}
\node at (4.5,-3.7) {$\longrightarrow^+_{in}$};
\node[align=center] at (7,-3.1) {{\tt $seqt($nth\_index$(l)(1) + 1)|_{l} = $} \\ {\tt $seqt(3)$ }};
\drawHterm{blue}{.7}{30}{north west lines}{blue}{7,-4.1}
\node at (9.5,-3.7) {$\longrightarrow^+_{in}$};
\node[align=center] at (12,-3.1) {{\tt $seqt($nth\_index$(l)(2) + 1)|_{l} = $} \\ {\tt $seqt(m)$ }};
\drawHterm{blue}{.7}{30}{crosshatch}{blue}{12,-4.1}
\node at (14.5,-3.7) {$\longrightarrow^+_{in} \cdots$};

\node at (2.5,-4.5) {An infinite innermost reduction sequence from $seqt(0)|_l$.};

\end{scriptsize}
\end{tikzpicture}
	
\begin{minipage}{.98\textwidth}\begin{scriptsize}
		\begin{tikzpicture}[scale=0.72, transform shape]
			\draw (-3.5,.5) rectangle (13.25,-2.3); 
			\draw[red] (-3.2,.4) rectangle (1.7,-2.2); 
			
			\node[align=center] at (-2,0) {$s$};
			\drawterm{black}{1.5}{70}{white}{-2,-1.7}
			\drawterm{red}{.7}{30}{white}{-2.5,-2}
			\drawterm{blue}{.7}{30}{white}{-2,-2}
			\drawterm{Green}{.7}{30}{white}{-1.5,-2}
			\node at (-.8,-1) {$\rightarrow^*_{in_{> \lambda}}$};
			
			\node[align=center] at (0.5,0) {$seqt(0)$};
			\drawterm{black}{1.5}{70}{white}{0.5,-1.7}
			\drawterm{red}{.7}{30}{white}{0,-2}
			\drawHterm{red}{.7}{30}{north east lines}{red}{0,-2}
			\drawterm{blue}{.7}{30}{white}{.5,-2}
			\drawHterm{blue}{.7}{30}{north east lines}{blue}{.5,-2}
			\drawterm{Green}{.7}{30}{white}{1,-2}
			\drawHterm{Green}{.7}{30}{crosshatch dots}{Green}{1,-2}
			
			
			\draw[red] (2.3,0.2) rectangle (5.6,-1.2); 
			\draw[thick] (2.1,0.4) rectangle (12.7,-2.2); 
			
			\node at (3,0) {$s|_l$};
			\drawterm{blue}{.7}{30}{white}{3,-1}
			\node at (4,-.7) {$\rightarrow^*_{i}$};
			\node at (5,0) {$seqt(0)|_l$};
			\drawHterm{blue}{.7}{30}{north east lines}{blue}{5,-1}

			\node at (6.2,-.7) {$\rightarrow_{i}$};
			\node at (7,0) {$seqt(2)|_l$};
			\drawHterm{blue}{.7}{30}{crosshatch dots}{blue}{7,-1}
			\node at (8,-.7) {$\rightarrow_{i}$};
			\node at (9,0) {$seqt(3)|_l$};
			\drawHterm{blue}{.7}{30}{north west lines}{blue}{9,-1}
			\node at (10,-.7) {$\rightarrow_{i}$};
			\node at (11,0) {$seqt(m)|_l$};
			\drawHterm{blue}{.7}{30}{crosshatch}{blue}{11,-1}
			\node at (12,-.7) {$\rightarrow_{i}\cdots$};
			\node at (6.5,-1.8) {The infinite innermost reduction sequence from subterm $s|_l$.};

		\end{tikzpicture}
	\end{scriptsize}
\end{minipage}
	
	\caption{Proof intuition: building an infinite innermost derivation of an argument $l$ as concatenation of a finite  and an infinite non-root innermost derivation of terms.}\label{fig:mintNRnoetherian}
\end{figure}

This proof follows by contraposition, by assuming the non noetherianity of the $\underset{s}{\rightarrow} \,_{in_{>\lambda}}$ relation and building then an infinite derivation for some argument of $s$, as illustrated in Figure \ref{fig:mintNRnoetherian}. Thus, initially one would have an infinite sequence $seqt$ of descendants of term $s$ where each one is related to the next one by one step of non-root reduction. From this sequence, since there is a finite number of possible arguments where the reductions can take place and infinitely many reductions taking place in non-root positions, i.e., argument positions, one uses the pigeonhole principle to ensure that there exists some argument position $l$ that satisfies the predicate {\tt inf\_red\_arg\_in\_inf\_nr\_im\_red}. This allows the use of function {\tt nth\_index} to extract exactly the index of the sequence where such reduction occurs. Then the required infinite derivation is built in two steps. First, since one has, by definition, that $s \overset{>\lambda}{\rightarrow} seqt(0)$, this leads to a finite sequence of reduced terms that will be used. Given that every argument of a term innermost reduces at root position to the argument of a reduced term by non-root reductions (lemma {\tt non\_root\_rtc\_reduction\_of\_argument} in theory {\tt innermost\_reduction.pvs}), the subterms of each element of this derivation at the chosen argument position is used to the first portion of the infinite sequence. Finally, the function {\tt nth\_index} is used to extract from sequence $seqt$ those indices where reductions occur in the selected argument, keeping this argument intact whenever the reduction does not occur in such indices (result given in lemma {\tt argument\_protected\_in\_non\_nth\_index}). Then, for each term obtained by a reduction on the $l$-th argument on this (now infinite) derivation, its subterm at argument $l$ is used to build the second and final portion of the infinite sequence. 

\subsubsection{Construction of Non-root Innermost Normal Forms for \emph{mint} terms}

Since a \emph{mint} term $s$ is noetherian regarding $\underset{s}{\rightarrow} \,_{in_{>\lambda}}$, as previously shown, in an infinite derivation starting from $s$ there  exists an index where the first innermost reduction in the root position occurs. This result is formalized in lemma {\tt inf\_inn\_deriv\_of\_mint\_has\_min\_root\_reduction\_index}.

\hspace{-4mm}\begin{minipage}{.98\textwidth}\begin{footnotesize}
	\begin{lstlisting}[label = {spec:hasMinRootRed}, caption = {Lemma stating the obligation of a first root reduction on infinite innermost derivations.}, mathescape, frame = single]
  inf_inn_deriv_of_mint_has_min_root_reduction_index: LEMMA
   $\forall(seq$: sequence[term]):
     $(\mnt_{i}(seq(0)) \;\;\land\;\; \forall (i:$nat): innermost_reduction?$(E)(seq(i), seq(i+1))) \rightarrow $
       $\exists (j:nat): seq(j) \overset{\lambda}{\rightarrow}_{i} seq(j+1) \;\;\land$ 
       $\forall (k:nat): seq(k)\overset{\lambda}{\rightarrow}_{i}  seq(k+1) \rightarrow k >= j $
	\end{lstlisting}
\end{footnotesize}
\end{minipage}

This lemma is formalized by providing as the first index required the minimum index of the infinite derivation where the reduction takes place at root position. The function minimum ({\tt min}) of PVS, just as function {\tt choose}, also requires a proof of non-emptiness of the set used as parameter. With the noetherianity provided by lemma {\tt mint\_is\_nr\_inn\_terminating}, this non-emptiness constrain is obtained through an auxiliary result over noetherian relations restricted to an initial element that are subsets of some non noetherian relation, which is given by lemma {\tt non\_noetherian\_and\_noetherian\_rest\_subset} in the {\tt restricted\_reduction.pvs} theory. This lemma provides an index of this infinite derivation where the given relation, i.e., $\underset{s}{\rightarrow} \,_{in_{>\lambda}}$ does not hold. 

Notice that, until this point, some infinite reduction sequence is being considered in the proof. However, the DPs are not extracted from the whole terms in this derivation. Instead, a \emph{mint} term is innermost reduced until reaching an innermost normal form and then the rule applied to the root builds the DP. Thus, at this point, the extraction of the DP would be possible. But since the instance of this DPs is crucial for building an infinite chain, it is important to know that not only the term that initiated the infinite derivation will be at some point reduced at root position, but which exact term was reached before such reduction. 

In order to be able to extract the DP and substitution required to proceed with the proof, one obtains finally that every \emph{mint} term non-root innermost derives into a term that has its arguments in normal form.

\hspace{-4mm}\begin{minipage}{.98\textwidth}\begin{footnotesize}
	\begin{lstlisting}[label = {spec:ExistsMintToNRNF}, caption = {Lemma for obtaining a non-root normal form term from a \emph{mint} term.}, mathescape, frame = single]

  mint_reduces_to_int_nrnf_term: LEMMA
    $\forall(s | \mnt_{i}(s)):
      \exists(t | \uparrow_{i}(t)): 
        s \rightarrow^*_{in_{>\lambda}} t \;\; \land \;\; nf_{\overset{>\lambda}{\rightarrow}}(t)$
	\end{lstlisting}
\end{footnotesize}
\end{minipage}

The proof follows as an application of previous lemma, choosing the term at the index where the first reduction at root position takes place, since this term is in innermost normal form. Indeed, this term will be a normal instance of the \lhs\ of some rule.

\subsection{Existence of DPs}\label{ssec:ExistenceOfDP}

The term obtained in previous subsection is an innermost non-terminating term such that it is also non-root innermost normalized.  Such non-root normalized terms should innermost reduce at root position, see $\overset{\lambda}{\rightarrow}_{i}$-reductions in Figure \ref{fig:infDerivInfDP}. These reductions from vertically to diagonally striped triangles give rise to the desired DPs. An important observation is that such terms reduce at root position with a rule and a normal substitution. The substitution should be normal since the terms are non-root innermost normal forms.

The following key auxiliary lemma provides the important result that such  normal instances of \rhs's of rules applied as before and that have minimal innermost non-terminating  subterms give rise to dependency pairs.  The innermost non terminality of the terms will guarantee the existence of such subterms.

	\hspace{-4mm}\begin{minipage}{.98\textwidth}\begin{footnotesize}
	\begin{lstlisting}[label = {spec:obtainingDPs}, caption = {Lemma for obtaining the desired DP from a \emph{mint} term with normal substitution.}, mathescape, frame = single]
  normal_inst_of_rule_with_mint_on_rhs_gives_dp_alt: LEMMA 
    $\forall(e \in  E, \sigma:($normal_sub?$(E)), 
\pi \in Pos(\rhs(e)\sigma)):$
      $\mnt_{i}(\rhs(e)\sigma|_{\pi}) \rightarrow $dep_pair_alt?$(E)(e,\pi)$
	\end{lstlisting}
\end{footnotesize}
\end{minipage}

The proof only requires showing that $\rhs(e)|_{\pi}$ is defined. For this, initially it must be ensured that $\pi$ is indeed a non variable position of $\rhs(e)$. But $\sigma$ is normal, thus, since the premise  $\mnt_{i}(\rhs(e)\sigma|_{\pi})$ implies innermost reducibility of $\rhs(e)\sigma|_{\pi}$, if $\pi$ were a variable position or a position introduced by this substitution, there would be a contradiction to its normality. This result is formalized separately in lemma {\tt reducible\_position\_of\_normal\_inst\_is\_app\_pos\_of\_term} that states that reducible subterms of normal instances of terms appear only at non variable positions of the original term. Then, by the main result of the last subsection, i.e., lemma {\tt mint\_reduces\_to\_int\_nrnf\_term}, one has that $rhs(e)\sigma|_{\pi} \rightarrow^*_{in_{>\lambda}} t$ for some term $t$ such that $\mnt_{i}(t)$ and $nf_{\overset{>\lambda}{\rightarrow}}(t)$. Then, the term $t$ has a defined symbol on its root. Thus, it only remains to prove that the root symbol of $rhs(e)\sigma|_{\pi}$ and $t$ is the same, which is an auxiliary result formalized by induction on the length of the non-root (innermost) derivation in corollary {\tt non\_root\_ir\_preserves\_root\_symbol} for non-root innermost derivations.

\subsection{Construction of Chained DPs}\label{ssec:BuildingDPsSubs}

So far the existence of the elements needed for the proof was formalized. Now, one builds in fact the elements as in Figure \ref{fig:infDerivInfDP}. Initially, a \emph{mint} term is non-root innermost normalized through the function {\tt mint\_to\_int\_nrnf} in Specification \ref{spec:mintToNRNF}. The existential result given by the lemma in Specification \ref{spec:ExistsMintToNRNF} on subsection \ref{ssec:InNormalFromMint} allows the use of the PVS {\tt choose} operator. 

\hspace{-4mm}\begin{minipage}{.98\textwidth}\begin{footnotesize}
	\begin{lstlisting}[label = {spec:mintToNRNF}, caption = {Function to provide an innermost non-terminating non-root normal form term from a \emph{mint} term.}, mathescape, frame = single]
  mint_to_int_nrnf$(E)(s: term | \mnt_{i}(s)):$ term = 
    choose$(\{t: $term$ | s \rightarrow^*_{in_{>\lambda}} t) \; \land \;  nf_{\overset{>\lambda}{\rightarrow}}(t) \; \land \; \uparrow_{i}(t)\})$
	\end{lstlisting}
\end{footnotesize}
\end{minipage}

Since this new non-root innermost normalized term is also innermost non-terminating, there exists some rule and normal substitution for allowing innermost reduction of this term at its root. Furthermore, the term obtained from this reduction will be also innermost non-terminating, i.e., it will have a \emph{mint} subterm at some position of the \rhs\ of the used rule. This property is formalized in lemma {\tt reduced\_nit\_nrnf\_has\_mint} specified as in Specification \ref{spec:NRNFredHasMint}. 

\hspace{-4mm}\begin{minipage}{.98\textwidth}\begin{footnotesize}
	\begin{lstlisting}[label = {spec:NRNFredHasMint}, caption = {Lemma ensuring the existence of \emph{mint} terms on reductions of innermost non-terminating non-root normal form.}, mathescape, frame = single]
  reduced_nit_nrnf_has_mint: LEMMA
    $\forall(s: $term$ | \mnt_{i}(E)(s)):$
      $\exists(\sigma: $Sub$, e: $rewrite_rule$ \;|\; e \in E, \pi \in Pos(\rhs(e))):$
        $lhs(e)\sigma = $mint_to_int_nrnf$(E)(s) \;\land\;  \mnt_{i}(\rhs(e))\sigma|_{\pi})$
	\end{lstlisting}
\end{footnotesize}
\end{minipage}

This lemma is formalized applying the existential results of Subsection \ref{ssec:ExistenceOfDP} for obtaining the normal substitution $\sigma$ and the rule $e$ and, the results of the Subsection \ref{ssec:ExistsMint} to obtain a position $\pi$ such that   $\mnt_{i}(\rhs(e))\sigma|_{\pi})$.

Lemma {\tt reduced\_nit\_nrnf\_has\_mint} allows one to use  {\tt choose} to pick the rule and position leading to the DP and the substitution that will allow chaining the DP with the next DP originated from the \emph{mint} term $\rhs(e)\sigma|_\pi$ as specified in function {\tt dp\_and\_sub\_from\_int\_nrnf} given in Specification \ref{spec:obtainingDPandSub}. Here it is clear why this construction is facilitated by the use of the alternative definition of DPs that includes both the rule and the position. 

\hspace{-4mm}\begin{minipage}{.98\textwidth}\begin{footnotesize}
	\begin{lstlisting}[label = {spec:obtainingDPandSub}, caption = {Function to obtain the desired DP and substitution.}, mathescape, frame = single]
 dp_and_sub_from_int_nrnf$(E)(s: $term$ | \mnt_{i}(s)):$ [dep_pair_alt$(E)$, Sub] = 
  LET $sub\_e\_p = $choose$(\{(\sigma: $Sub$, e \in E, \pi \in Pos(\rhs(e))) \;| $
                     $\lhs(e)\sigma =$ mint_to_int_nrnf$(E)(s),         \mnt_{i}(\rhs(e))\sigma|_{\pi})\})$ 
  IN $((sub\_e\_p'2,sub\_e\_p'3),sub\_e\_p'1)$
\end{lstlisting}
\end{footnotesize}
\end{minipage}

Whenever this function has as input a term that is an instance of the \rhs\ of a DP that is in non-root innermost normal form, the resulting DP and substitution will be chained with the DP and substitution used to build the input term. This result is specified in lemma {\tt next\_inst\_dp\_is\_inn\_chained\_and\_mnt} given in Specification\ref{spec:ObtentionOFnextInstDP}, where the desired alternative DPs are transformed into standard DPs in order to allow the analysis through the predicate {\tt inn\_chained\_dp?}:

\hspace{-4mm}\begin{minipage}{.98\textwidth}\begin{footnotesize}
	\begin{lstlisting}[label = {spec:ObtentionOFnextInstDP}, caption = {Lemma ensuring that the obtained DPs and substitutions are chained.}, mathescape, frame = single]
next_inst_dp_is_inn_chained_and_mnt: LEMMA
  $\forall (E)(\;dp:$ dep_pair_alt$(E),$
       $\sigma:$Sub$ \;|\; \mnt_{i}(\rhs(dp'1)\sigma|_{dp'2})) \;\;\land\;\; nf_{\overset{>\lambda}{\rightarrow}}(\lhs(dp'1)\sigma)\;\;)\;\;:$
     LET $std\_dp = (\lhs(dp'1), \rhs(dp'1)|_{dp'2})$,
         $next\_dp\_sub$ =  dp_and_sub_from_int_nrnf$(E)(\rhs(dp'1)\sigma|_{dp'2})$,
         $next\_std\_dp= (\lhs(next\_dp\_sub'1'1),\rhs(next\_dp\_sub'1'1)|_{next\_dp\_sub'1'2}), $
         $\sigma' = next\_dp\_sub'2$ IN
      inn_chained_dp?$(E)(std\_dp, next\_std\_dp)(\sigma, \sigma') \;\;\land\;\;  \mnt_{i}((next\_std\_dp'2)\sigma')$
  	\end{lstlisting}
\end{footnotesize}
\end{minipage}

The formalization of this lemma is quite simple in its core. However, since transformations between the standard and alternative notions of DPs are used, the proof of some typing conditions are required in order to ensure  type correctness. Once circumvented the typing issues, one must only guarantee the innermost chained property for the input DP and substitution and the resulting DP and substitution created and that the instantiated subterm of the \rhs\ of the new DP is a \emph{mint} term. Notice that the latter property is a direct result of the type of the PVS {\tt choose} operator  used in function {\tt dp\_and\_sub\_from\_int\_nrnf}; indeed, this property was included (and formalized) as part of this lemma just to avoid needing to repeatedly ensure non-emptiness of the used set, since this result is used several times throughout the rest of the formalization. To guarantee that the DPs are chained is also straightforward, since {\tt dp\_and\_sub\_from\_int\_nrnf} is defined over {\tt mint\_to\_int\_nrnf}, which gives a term with type as a non-root innermost normal form of the \emph{mint} input, i.e., exactly the definition given by predicate {\tt inn\_chainned\_dp?}; using notation of the lemma: $\rhs(dp'1)|_{dp'2}\sigma \rightarrow^*_{in_{>\lambda}}$ $\lhs(next\_dp\_sub'1'1)\sigma'$.

This result allows the specification of a function using predicate subtyping, a very interesting feature available in PVS. Using this feature, elaborate predicate types can be assigned to the outputs of  functions, and type checking will automatically generate the \emph{type check conditions} (TCCs) to ensure well-definedness of the function.  Although used in other functions through the formalization, the most interesting application of this feature happens in the next function that outputs a pair  for an input pair of DP and substitution, and where the type of the output uses the predicates {\tt in\_chained\_dp?}  and  $\mnt_{i}$.  The generated TCCs are not proved automatically; however, to ensure that the type predicates hold, typing provided in the lemma {\tt} given in Specification \ref{spec:ObtentionOFnextInstDP} \sout{previous lemma} are applied. 
 
 \hspace{-4mm}\begin{minipage}{.98\textwidth}\begin{footnotesize}
 	\begin{lstlisting}[label = {spec:nextDPsub}, caption = {Function to obtain adequate next DP and substitution.}, mathescape, frame = single]
next_dp_and_sub$(E)(\;\;dp:\;$dep_pair_alt$(E),$
$\mbox{\hspace{4cm}} \sigma:\;$Sub$ \;|\;  \mnt_{i}(rhs(dp'1)\sigma|_{dp'2})\; \land nf_{\overset{>\lambda}{\rightarrow}}(lhs(dp'1)\sigma)\;\;)\; :$
  {$\;(next\_dp:$ dep_pair_alt$(E), $
    $next\_\sigma:\;$Sub) | inn_chained_dp?$(E)(dp, next\_dp)(\sigma, next\_\sigma)\; \land$
      $\mbox{}\hspace{2cm}\mnt_{i}(rhs(next\_dp'1)next\_\sigma|_{ next\_dp'2}))\;$} = 
 	 dp_and_sub_from_int_nrnf$(E)(rhs(dp'1)\sigma)|_{dp'2})$
   	\end{lstlisting}
\end{footnotesize}
\end{minipage}
	
	Applying  {\tt dp\_and\_sub\_from\_int\_nrnf} (Specification \ref{spec:obtainingDPandSub}) to a \emph{mint} term built from a pair of  DP and substitution (in the way done in the body of the function {\tt next\_dp\_and\_sub}), one provides as output a pair of DP and substitution with the specified subtyping predicates, guaranteeing that the input and output are chained.

\subsection{Construction of the Infinite Innermost Dependency Chain}\label{ssec:BuldingInfInDepChain}

 With the possibility of creating new DPs and substitutions from \emph{mint} terms, it is possible to build, inductively, an infinite DP chain from any  innermost non-terminating term. However, PVS syntax makes this construction a little bit tricky, since its functional language only allows directly construction of lambda-style or recursive functions. A lambda-style function to create such infinite chain is not possible, since the construction of every pair of DP and substitution depends on the previous one in the chain. But a direct construction of a recursive function is also problematic since  the use of the {\tt choose} operator in several steps of this construction makes it difficult to guarantee its determinism and then its functionality. 
 
 A simple solution for this problem is to use the recursion theorem to provide the existence of a function from naturals to pairs of a DP and a substitution such that each pair generates the next pair in the chain according to the function {\tt next\_dp\_and\_sub}, implying that contiguous images are chained. 
 
 The recursion theorem is given in Specification \ref{spec:recursionTheorem}. It states that for all predicates $X$ over a set $T$, initial element $a$ in $X$ and function $f$ over elements of $X$, there exists a function $u$ from naturals to $X$ such that  the images of $u$ are given by the sequence $a, f(a), \ldots, f^n(a), \ldots$. 
 
 \hspace{-4mm}\begin{minipage}{.98\textwidth}\begin{footnotesize}
 	\begin{lstlisting}[label = {spec:recursionTheorem}, caption = {The recursion Theorem.}, mathescape, frame = single]
  recursion_theorem : THEOREM
    $\forall (X : $set$[T],  a \in X, f : [(X) -> (X)]):$
      $\exists(u: [nat -> (X)]): u(0) = a \land \forall(n: $nat$): u(n+1) = f(u(n))$
 	\end{lstlisting}
 \end{footnotesize}
\end{minipage}
 
 To use this theorem, the predicate is instantiated with pairs of DP and substitution of the type of the parameters of the function {\tt next\_dp\_and\_sub}, i.e., {\footnotesize$ (dp:\;${\tt dep\_pair\_alt}$(E),  \sigma:\;${\tt Sub} $ \;|\;  \mnt_{i}(rhs(dp'1)\sigma|_{dp'2})\; \land nf_{\overset{>\lambda}{\rightarrow}}(lhs(dp'1)\sigma))$}.

The first element of the sequence $a$ is instantiated as the pair of DP and substitution, obtained from the initial term starting any infinite innermost derivation, according to the techniques given is subsections \ref{ssec:ExistsMint}, \ref{ssec:InNormalFromMint} and \ref{ssec:ExistenceOfDP}.  As expected, the function from pairs to pairs is chosen as {\tt next\_dp\_and\_sub}.  The recursion theorem guarantees just the existence of a total function from naturals to the sequence inductively built using function {\tt next\_dp\_and\_sub} starting from the initial pair. But the choice of this function assures by its predicate subtyping that each pair of consecutive pairs are in fact chained.  

As a consequence of all that, the sufficiency lemma (\ref{spec:suffLemma}) is obtained. 

\hspace{-4mm}\begin{minipage}{.98\textwidth}\begin{footnotesize}
	\begin{lstlisting}[label = {spec:suffLemma}, caption = {The sufficiency lemma for DP termination.}, mathescape, frame = single, framexrightmargin=0cm]
  dp_termination_implies_noetherian: LEMMA 
    $\forall(E): $  inn_dp_termination?$(E) \rightarrow $noetherian?$(\rightarrow_{i})$
	\end{lstlisting}
\end{footnotesize}
\end{minipage}

\section{Related Work}\label{sec:RelatedWork}

There are several methods of semi-decision to address the analysis of termination, among them, the well-known \emph{Ranking functions} implemented in PVS as termination TCCs, as mentioned in the introduction.  A more recent criterion to verify termination of functional programs is the so-called \emph{size-change principle} (SCP, for short) \cite{LeeJonesBen01}. This principle does not require decreasingness after each recursive call, but strict decreasingness (using a measure regarding some well-founded order) for each possible infinite ``cycle'' of recursive calls; thus, if such a measure exists,  infinite computations are not possible since they will imply infinite decreasingness (over a well-founded order). The SCP and DPs criterion are compared in \cite{ThiemannGiesl05} taking into account termination, innermost termination and evaluation of functional specifications. One approach of the SCP is given by the technology of \emph{calling contexts graphs} (CCG, for short) \cite{ManoliosVroon06}, which implements the SCP by representing all possible executions of a functional program as paths in a graph in which nodes are labeled by the different occurrences of function calls in it. More precisely, each node corresponds to a so-called \emph{calling context} that consists of  the formal parameters of a function in which a function call is specified, the actual parameters of the function call, and the conditions that lead to the execution of the function call.  Possible computations are then characterized as sequences of calling contexts related to paths in that graph, and termination is analyzed regarding the behavior of measurements on the possible circuits in the CCG. 

	
Application of the DP termination criterion is recurrent in termination tools. Expressive developments, which have been implemented in such tools, include the work of Alarcon and Lucas \cite{alarconLucas09, alarconLucas10}  who succesfully applied  DPs for several strategies and restrictions for (context-sensitive) TRSs, and the work of Sternagel and Thiemann \cite{Thiemann10} who formalized correctness of the generalization of the DP criterion to Q-restricted TRSs and implemented DP  for checking termination of functional specifications. The latter work uses the methodology of translating functional specifications into TRSs that are then checked for termination by the DP criterion. The current work focuses on the formalization of correctness of the particular case of innermost DP as another termination criterion to be added to those available in the PVS theory  of the functional PVS0 specifications.

Formalizations of the theorem of soundness and completeness of DPs (DP theorem, for short) are available in several proof assistants. In \cite{blanqui11}, Blanqui and Koprowski described a formalization of the DP theorem for the ordinary or standard reduction relation that is part of the CoLoR library developed in Coq for certifying proofs of termination. The formalized result is the DP theorem for the standard reduction relation, and not for the innermost termination. The proof in \cite{blanqui11}, as the current formalization,  uses the non-root reduction relation (internal reduction) and the reduction at root position relation (head reduction). Instead of building infinite chains from infinite derivations, it assumes a well-founded relation over the set of chained DPs to conclude noetherianity of the standard reduction relation.
Also, the library Coccinelle \cite{contejean07} includes a formalization in Coq of DP theorem that defines a relation between instances of \lhs\ of DPs and proves the equivalence between well-foundedness of this relation and well-foundedness of the reduction relation of a given TRS.  To chain DPs instances of the lists of arguments of \lhs's and \rhs's of DPs, which are headed by the same function symbol, are related by the reflexive-transitive closure of the rewriting relation (avoiding in this way the use of tuple symbols). The formalization also considers a refinement of the notion of DPs, which avoids DPs generated by a rule, where the \rhs\ of the DP appears  also as a subterm of the \lhs\ of the rule.    

A formalization of the DP theorem for the standard reduction relation is also present in the proof assistant Isabelle, as part of the library for rewriting IsaFoR briefly described in  \cite{Thiemann10}. In this formalization the original signature of the TRS is extended with new tuple symbols for substituting the defined symbols (see comments after Definition \ref{def:depPairs} of DPs), which implies the analysis of additional properties of the new term rewriting system induced over the extended signature and also properties relating this new rewriting system with the original one.  The proof, as in the current formalization, builds an infinite chain from an infinite derivation and vice-versa. This work brings interesting features, such as the use of the same refinement of DPs as the formalization in Coccinelle and that it was done for a full definition of ``Q-restricted'' rewriting, providing in this manner a general result that has as corollaries both the DP theorem for the standard and the innermost reduction relations, the former given explicitly.  Essentially, for TRSs $E$ and $Q$, the $Q$-restricted relation, denoted as $\overset{Q}{\rightarrow}_E$, is defined as the relation such that $s\overset{Q}{\rightarrow}_E t$ iff $s \rightarrow_E t$ at some position $\pi$ such that proper subterms of $s|_\pi$ are normal regarding $Q$; so $\overset{\emptyset}{\rightarrow}_E$ and $\overset{E}{\rightarrow}_E$ correspond respectively to the standard and the innermost reduction relations \cite{Giesl05}.  This formalization is used to provide a sound environment to certify concrete termination proofs in an automatic way by the tool CeTA \cite{certTermCeta09}.     Formalization of the DP criterion for the  ordinary rewriting relation is also included in the PVS theory {\tt TRS} (and was done as part of this job), but as mentioned in the introduction, the emphasis  in this work is on the innermost case since it is the one related to the operational semantics of functional specifications.

\section{Relating TRS Termination to Functional Program Termination}\label{ssec:DPvsFT}
	The CCGs technology has the advantage of allowing combinations of a finite family of measures at each node of a possible circuit, simplifying in this manner the formulation of a single and complex measure that works (decreases) for all possible circuits. These combinations are also implemented in the so-called Matrix Weighted Graphs (MWG) developed by Avelar in \cite{Andreia2014}. All these technologies (TCC, SCP, CCG, MWG) to verify termination are implemented and formalized to be equivalent in the PVS library {\tt PVS0}. This theory uses a 
	simple functional language also called PVS0, used to reason about termination of PVS programs while simplifying proofs. Expressions of PVS0 programs are described by the following grammar.
	\[\Expr ::=  \cnst\ |\ \vr\ |\ op1(\Expr )\ |\ op2(\Expr ,\Expr)\ |\ \rec(\Expr )\ |\ \ite(\Expr,\Expr,\Expr )\]
	
	This grammar is specified  as an abstract datatype that allows one to have  unary and  binary operators, which are interpreted separately as built-in operators (\cite{Ramos18}). Then it is possible to use elements of this datatype, such as its constructors, accessors and recognizers in the proof process regarding these expressions. Furthermore, in order to provide all elements required to specify a program in the PVS0 language, each program requires lists of the interpretation of its unary and binary operators ($\OPa$ and $\OPb$), a fixed constant to be the false value ($\False$) and a expression representing the body of the function for the program itself. This quadruple is called a PVS0 program ($\pvso$).
	
	Formalizations relate the operational semantics of the PVS0 language and termination criteria.  The semantics of termination for an expression $e$ is given by two different operators. The first one is a predicate and the second one a function. In both cases the  evaluation of $e$ depends on a given $\pvso$ program, i.e., it depends on the interpretation of lists of unary and binary operators ($\OPa$ and $\OPb$), the false value ($\False$), and the expression representing the function (program itself) where the evaluation must take place $e_f$.  Input and output values  will be denoted as $v_i$ and $v_o$, respectively. The semantic evaluation predicate is given by $\varepsilon$ (Table \ref{tab:varepsilon}), and the intuition is that the program $\pvso$ evaluates the expression $e$ with input $v_i$ as $v_o$. 
	
\begin{table}[!ht]	
\caption{Semantic evaluation predicate $\varepsilon$}
\label{tab:varepsilon}
	\[\begin{array}{rcl}
	\varepsilon(\OPa,\OPb,\False,e_f) (e, v_i, v_o) & := & \CASES e\, \OF\\
	&& \hspace{-3cm}
	\begin{array}{rl}
	\cnst(v) \; : \; & v_o = v; \\ \vr \; : \; & v_o = v_i; \\
	\opa(j,e_{1}) \; : \; & j < |\OPa| \;\land
	\exists\, v' \in \Val:\\
	& \varepsilon(\OPa,\OPb,\False,e_f)(e_{1},  v_i,
	v') \land  v_o = \OPa(j)(v');\\ 
	\opb(j,e_{1}, e_{2})  \; : \; &j < |\OPb| \;\land \exists\, v',v'' \in \Val: \\
	& \varepsilon(\OPa,\OPb,\False,e_f)(e_{1},v_i, v') \ \land \\ 
	& \varepsilon(\OPa,\OPb,\False,e_f)(e_{2},  v_i, v'')\ \land \\  
	& v_o = \OPb(j)(v', v'');\\
	\rec(e_{1}) \; : \; & \exists\, v' \in {\cal V}al:
	\varepsilon(\OPa,\OPb,\False,e_f)(e_{1},  v_i,
	v')\ \land \\  
	&\varepsilon(\OPa,\OPb,\False,e_f)( e_f,v', v_o)\\
	\ite(e_{1}, e_{2}, e_{3}) \; : \; & \exists\, v'
	:\varepsilon(\OPa,\OPb,\False,e_f)(e_{1},
	v_i, v')\; \land \\  
	&\IF v' \neq \bot\;\THEN\\
	& \quad \varepsilon(\OPa,\OPb,\False,e_f)(e_{2},  v_i, v_o) \\
	& \ELSE \\ & \quad \varepsilon(\OPa,\OPb,\False,e_f)(e_{3},  v_i, v_o).
	\end{array} 
	\end{array}\] 
	\end{table}
	
	The first  semantic	termination notion for a $\pvso$ program is then given as: 
	\[T_{\varepsilon}(\pvso) := \forall\, (v \in \Val): \exists\, ({v_o} \in \Val) \,:\,
	\varepsilon(\pvso)(\pvso_e,{v_i},{v_o}).\]
	which holds for a given program $\pvso$ whenever, for every input 
	$v_i$, the evaluation of the program expression $\pvso_e$ on the value $v_i$  holds for some output value $v_o$. 
	
	The second specification for semantic evaluation is given as the recursive function $\chi$ in Table \ref{tab:chi}. This function, in addition to the PVS0 program $\pvso$, the input expression $e$ and the input value $v_i$, has a parameter $n$ that is a natural number giving the maximum allowed number of nested recursive calls. This function returns an output value whenever it is possible to evaluate it allowing at most $n$ nested recursive calls and a ``none''  value ($\noval$) otherwise.
	
	\begin{table}[!ht]
	\caption{Semantic evaluation function $\chi$}
\label{tab:chi}
	\[\begin{array}{rcl}
	\chi(\OPa,\OPb,\False,e_f) (e, v_i,n) & := &  \IF n = 0 \;\THEN
	\noval \;\ELSE
	\;\CASES e \;\OF\\
	&& \hspace{-3cm}
	\begin{array}{rl}
	\cnst(v) \; : \; & v;\\ 
	\vr \; : \; & v_i; \\
	\opa(j,e_{1}) \; : \; & \IF j < |\OPa| \;\THEN\\
	& \ \ \ \LET v' = \chi(\OPa,\OPb,\False,e_f)(e_{1}, v_i,n) \; \IN \\
	&\ \ \ \IF v' = \noval \;\THEN \noval \;\\&\ELSE \OPa(j)(v')\\ 
	&\ELSE\noval;\\ 
	\opb(j,e_{1}, e_{2})  \; : \; &\IF j < |\OPb| \THEN\\
	& \ \ \ \LET v'\ =\chi(\OPa,\OPb,\False,e_f)(e_{1},v_i,n),\\ 
	& \ \ \ \ \ \ \ \ \ v''=\chi(\OPa,\OPb,\False,e_f)(e_{2}, v_i,n)\; \IN \\  
	& \ \ \ \IF v' = \noval \;\lor v'' = \noval \;\THEN \noval \;\\
	& \ \ \ \ELSE \OPb(j)(v', v'')\\
	&\ELSE\noval;\\
	\rec(e_{1}) \; : \; & \LET v' = \chi(\OPa,\OPb,\False,e_f)(e_{1}, v_i,n)\;\IN\\
	&\IF v' = \noval \;\THEN \noval \;\\ &\ELSE \chi(\OPa,\OPb,\False,e_f)(e_f,v',n-1);\\
	\ite(e_{1}, e_{2}, e_{3}) \; : \; & \LET v' = \chi(\OPa,\OPb,\False,e_f)(e_{1},v_i,n)\; \IN \\
	&\IF v' = \noval \;\THEN \noval \\
	&\ELSIF v' \neq \bot\;\THEN
	\chi(\OPa,\OPb,\False,e_f)(e_{2},v_i,n) \\
	& \ELSE \chi(\OPa,\OPb,\False,e_f)(e_{3},v_i,n).
	\end{array} 
	\end{array}\] 
	\end{table}
	
	Thus, the second notion of semantic termination is specified as the existence of a number of nested recursive calls allowing the evaluation of some value different from ``none'':
	\[T_{\chi}(\pvso) := \forall\, (v \in \Val) \,:\,\exists\, (n \in \nat) \,:\,
	\chi(\pvso)(\pvso_e,v,n) \neq \noval.\]
	
	These semantic termination specifications were formalized to be equivalent and used to formalize the correction and equivalence of the other mentioned termination criteria, namely, SCP, CCG, MWG and TCC termination criteria.  These formalizations are present in the {\tt PVS0} and {\tt CCG} libraries.
	
	Although the innermost DP termination criterion is formally related to noetherianity of the relation of chained DPs to verify termination of TRSs, it is  also adequate to reasoning about termination of functional programs under eager evaluation. The {\tt PVS0} theory also includes formalizations for this criterion for functional programs, which given the necessary adaptations, are at its core closely related to the specification given for CCGs. 
	
	In the libraries {\tt PVS0} and {\tt CCG} theories the concept of \emph{calling context} is specified as a triple capturing the information of a recursive call in a program as in \cite{ManoliosVroon06}:   the formal and actual parameters and the condition that leads to the recursive call.  Since PVS0 programs consist of a unique function, only the formal and actual parameters are required.  As an example consider the specification in PVS0 of the Ackermann function below.  Notice that the parameters are of type $\nat \times \nat$, $\bot=(0,0)$, $\top=(1,0) = \neg \bot$, and $a$ is the PVS0 program $(\OPa,\OPb,\bot,e_a)$.  The formal parameter is  $\vr \in \nat \times \nat$  encoding the two inputs of the Ackermann function and also the output of the function, which is given by the first component of the output, also of type $\nat \times \nat$.
	
	\[ \begin{array}{ll}
	\OPa(0)((m,n)) &:= \IF m = 0\;\THEN \top \;\ELSE \bot,\\[1mm]
	\OPa(1)((m,n)) &:= \IF n = 0\;\THEN \top \;\ELSE \bot,\\[1mm]
	\OPa(2)((m,n)) &:= (n+1,0),\\[1mm]
	\OPa(3)((m,n)) &:= \IF m > 0 \; \THEN (m-1,1)\;\ELSE \bot,\\[1mm]
	\OPa(4)((m,n)) &:= \IF n > 0 \; \THEN (m,n-1)\;\ELSE \bot,\\[1mm]
	\OPb(0)((m,n),(i,j)) &:= \IF m > 0 \; \THEN
	(m-1,i)\;\ELSE \bot,\\[1mm]
	e_a &:=  \ite(\opa(0,\vr), \opa(2,\vr),\\[1mm]
	&\ \ \ \ \ite(\opa(1,\vr),\rec(\opa(3,\vr)),\\[1mm]
	&\ \ \ \ \rec(\opb(0,\vr,\rec(\opa(4,\vr)))))).
	\end{array}\]
	
	Simplifying PVS0 notation,	the calling contexts to this PVS0 program are:
	
	
	
	

	\[ \begin{array}{l}
	\langle (m,n), m>0 \wedge n =0 , (m-1, 1)\rangle\\[1mm]
	\langle (m,n),  m>0 \wedge n>0 , (m-1, rec((m,n-1)))\rangle\\[1mm]
	\langle (m,n),  m>0 \wedge n> 0, (m, n-1)\rangle
	\end{array}\]
	
	In the specification conditions and actual parameters of the calling contexts are built from the formal parameter $(m,n)$ (that may be omitted) and the position of each recursive call by using expressions of the original signature. For instance, the condition and actual parameters of the first calling context above are given respectively as $\neg\opa(0,(m,n))\wedge\opa(1,(m,n))$ and $\opa(3,(m,n))$.
	
	Translating the functional program to a corresponding TRS as done in \cite{Sternagel10} is possible, but what should be essentially considered is the correspondence between the calling contexts of a PVS0 functional program and the DPs of an associated TRS.  Establishing such correspondence is enough since it allows the use of the associated calling contexts as a mechanism to check termination by DPs (or vice versa). For instance, if one consider the TRS and DPs for the Ackermann function as given in Examples \ref{ex:trsAck} and \ref{ex:dpAck}, whenever a pair of naturals $(m,n)$ \emph{matches}  $(s(x),0)$, exactly the condition of the first calling context holds:  $m>0\wedge n = 0$.  In addition, the actual parameter of the first calling context $(m-1, 1)$, \emph{matches} $(x, s(0))$. Similarly, this happens for the conditions and actual parameters of the second and third calling contexts.
	
	Since the analysis of termination using CCGs relies on sequences of values obtained by eager evaluation from the previous calling context  that must hold for the condition in the next context, the notion of innermost dependency chains is closely related to this usage.  The formalization of the equivalence of the relations of termination by CCGs and by DPs requires manipulation of different signatures for the associated functional programs and TRSs,  and requires proper association of evaluation of the conditions in the calling contexts and matchings of \lhs\ of rewriting rules that generate DPs. Furthermore, relating  chains of DPs and paths of calling contexts in a CCG also requires the adequate association of the two different signatures involved, in such a form that the ``evaluation'' of the DPs through non-root innermost normalization must correspond to the eager evaluation of functional expressions.  All this is the subject of associated research relating the formalization of correctness of the innermost DPs criterion, given in this paper, and termination criteria formalized in the theories  {\tt PVS0} and {\tt CCG}.
	
	Table \ref{tab:equiv} summarizes the equivalence results between termination criteria that are formalized in the PVS theories {\tt CCG} and {\tt PVS0}. It is important to stress here that the notion of DP termination was specified for PVS0 functional programs based on its relation with CCGs, which allowed its equivalence with the CCG criterion for this language to be formalized in a simple manner. For using the results formalized for the TRS theory, the aforementioned correspondence results between calling contexts and DPs must be formalized.
	
	\begin{table}[!ht]
	\caption{Termination equivalences for PVS0 programs and where to find them.}\label{tab:equiv}
		\begin{scriptsize}
			\begin{center}
				\begin{tabular}{|c|c|c|c|}
					\hline \textbf{Proof} & \textbf{Lemma name} & \textbf{File} & \textbf{Theory} \\\hline 
					
					$T_{\varepsilon} \Rightarrow TCC$& \argp{terminates\_implies\_pvs0\_tcc}		& \argp{\bf measure\_termination} 	   & \argp{ PVS0}\\\hline
					$TCC \Rightarrow T_{\varepsilon}$& \argp{pvs0\_tcc\_implies\_terminates} 	& 
					\argp{\bf pvs0\_termination}    		  & \argp{ PVS0}\\\hline
					$T_{\chi} \Leftrightarrow T_{\varepsilon}$	 & \argp{eval\_expr\_terminates}	  & \argp{\bf pvs0\_expr} 			   			& \argp{ PVS0}\\\hline
					$SCP \Rightarrow TCC$	& \argp{scp\_implies\_pvs0\_tcc} 				 & \argp{\bf scp\_iff\_pvs0}					& \argp{ PVS0}\\\hline
					$TCC \Rightarrow SCP$	& \argp{pvs0\_tcc\_implies\_tcc} 				  & \argp{\bf scp\_iff\_pvs0}			 		 & \argp{ PVS0}\\\hline
					$SCP \Rightarrow CCG$   & \argp{scp\_implies\_ccg\_pvs0}				& \argp{\bf pvs0\_to\_ccg} 		 		 & \argp{ PVS0}\\\hline
					$TCC \Rightarrow CCG$  & \argp{pvs0\_tcc\_implies\_ccg} 				 & \argp{\bf pvs0\_to\_ccg} 		   		  & \argp{ PVS0}\\\hline
					$SCP \Rightarrow CCG$  & \argp{scp\_implies\_ccg\_termination} 	 & \argp{\bf scp\_to\_ccg}					  & \argp{ CCG}\\\hline
					$CCG \Rightarrow SCP$  & \argp{ccg\_termination\_implies\_scp} 	 & \argp{\bf ccg}									& \argp{ CCG} \\\hline
					$DP \Rightarrow SCP$ 	& \argp{dp\_termination\_implies\_dp\_scp}  & \argp{\bf dp\_to\_tcc}                     & \argp{ PVS0}\\\hline
					$MWG \Leftrightarrow CCG$ & \argp{mwg\_termination\_iff\_ccg\_termination} & \argp{\bf ccg\_to\_mwg} & \argp{CCG} \\\hline
					$DP \Rightarrow TCC$ 	& \argp{dp\_termination\_implies\_dp\_dec}	& 
					\argp{\bf dp\_to\_tcc}                      & \argp{ PVS0}\\\hline
					$TCC \Rightarrow DP$ 	& \argp{dp\_dec\_implies\_dp\_termination}  & \argp{\bf dp\_termination}                & \argp{ PVS0}\\\hline
				\end{tabular} 
			\end{center}
		\end{scriptsize}
	\end{table}

\section{Discussion and Future Work}\label{sec:FutureWork}

A formalization in PVS of the soundness and completeness of the  Dependency Pairs criterion for innermost termination of TRSs was presented. The formalization follows the lines of reasoning of proofs given in papers such as \cite{ArtsGiesl00}. 

The kernel of the formalization consists of 56 lemmas, 34 of these being TCCs. These results are available in the specification and formalization files {\tt inn\_dp\_termination.pvs} and {\tt .prf} that have size 18KB and 747KB, respectively. The basic notions regarding Dependency Pairs are separately specified in file {\tt dependency\_pairs.pvs}, which add 4kb to the size of the whole specification and give rise to 8 TCCs, adding 13kb to the size of the formalization.  For achieving the formalization, the {\tt TRS} library of PVS, was extended with theories {\tt innermost\_reduction} and {\tt restricted\_reduction}, which include 38 lemmas, of which 17 are TCCs. Both these theories add 10KB of specification and 451 KB of proofs. The proof of necessity (in theory {\tt inn\_dp\_termination.pvs}) required 11\% of the whole size of the formalization file, while sufficiency required 80\%. The remaining 9\% of the formalization file deals with basic properties of DPs, and a lemma relating innermost DP termination with noetherianity of the innermost chain relation. From the total size used in the proof  of sufficiency, the formalization was split approximately into 11\%, 59\%, 6\%, 20\% and 4\% for the tasks  presented in Section \ref{sec:Sufficiency}: 
Existence of \emph{mint} Subterms  (Subsection \ref{ssec:ExistsMint}),
Non-root Innermost Normalization of \emph{mint} Terms (\ref{ssec:InNormalFromMint}),
Existence of DPs (\ref{ssec:ExistenceOfDP}), 
Construction of  chained DPs (\ref{ssec:BuildingDPsSubs}),
Construction of the Infinite Innermost Dependency Chain (\ref{ssec:BuldingInfInDepChain}), respectively. As expected from the discussion in Section \ref{sec:Sufficiency}, the formalizations of normalization of \emph{mint} terms and constructions of chained DPs were the most elaborate and the ones that required the most space.

In order to formalize the DP theorem for the  ordinary rewriting relation, 
a similar reasoning to the one used for innermost reduction was followed, but the involved properties were not reused to prove each other.  Notice for instance that necessity for the ordinary reduction, i.e.,  {\tt noetherian?(reduction?(E))} implies {\tt dp\_termination?(E)}, cannot be applied to infer {\tt inn\_dp\_termination?(E)}, if one has {\tt noetherian?(innermost\_reduction?(E))}.  The required properties were developed explicitly for the ordinary reduction relation, as done so far for the formalization of necessity theorem.  The main difference happened in the formalization of sufficiency, when  an infinite chain was built from an infinite derivation. Specifically, for the  innermost case, \emph{mint} terms are normalized regarding the non-root innermost relation, giving rise to a term that has an innermost reduction redex at its root (vertically striped small triangles in Figure \ref{fig:infDerivInfDP}),  while for the ordinary relation, the unique guarantee is that \emph{mnt} terms reduce at non-root positions into a term that can be reduced at its root position.  This small difference requires a few  adjustments in order to apply the rules on root position leading to the DPs that will produce the chain. The existence of DPs from such (non necessarily non-root normalized) terms follows from an argument based on the fact that the \emph{mnt} term starting the non-root derivation is non-root terminating. Thus, when a given rule is applied at root position of some of its non-terminating descendants, the substitution allowing the application of such rule may not have non-root nonterminating redexes. Other than that, the {\tt chained\_dp?} property also follows directly from the type of the chosen descendant term of a \emph{mnt} term where the first root reduction takes place.
The specification and formalization of correction of the DP criterion for ordinary rewriting is available in the theories {\tt dp\_termination.pvs} and {\tt .prf} and consist of 55 lemmas, 34 of which being TCCs.

In order to have a full formalization of the relation between the results presented in this work and the termination criteria formalized for PVS0 programs, it would be necessary to formalize the relation between the notions that analyze recursive calls leading to (possibly) infinite evaluations/innermost reductions. This requires not only dealing with different signatures for FPs and TRS, but also specifying the notion of ``reduction'' for rewriting terms with ``values", which must be ``evaluated'' into values as is the case of eager evaluation of functional programs for given input values.  In order to provide such a notion of evaluation of TRSs, constructor and defined symbols should be mapped into values when their innermost reduction for given expressions can lead to values. This result will allow linking all (formalized equivalent) criteria available to PVS0 functional programs and innermost DP criterion over TRSs.


\bibliography{jar.bib}

\end{document}